\documentclass{emulateapjBLAST}

\usepackage{url}
\usepackage{multirow}
\usepackage{natbib}
\usepackage{amsmath}
\slugcomment{To appear in The Astrophysical Journal}

\shorttitle{BLAST Observations of Resolved Galaxies}
\shortauthors{Wiebe et al.}

\newcommand{\IRAS}{{\it IRAS\/}}
\newcommand{\ROSAT}{{\it ROSAT\/}}
\newcommand{\ISO}{{\it ISO\/}}
\newcommand{\Spitzer}{{\it Spitzer\/}}
\newcommand{\Herschel}{{\it Herschel\/}}

\newcommand{\SANEPIC}{{\sc sanepic\/}}
\newcommand{\Almagest}{{\sc almagest\/}}
\newcommand{\SCUBATOO}{\hbox{SCUBA-2}}

\newcommand\dg{$^\circ$}
\newcommand\sqdg{deg$^2$}
\newcommand\hour{$^{\mathrm h}$}
\newcommand\minute{$^{\mathrm m}$}
\newcommand\second{$^{\mathrm s}$}
\newcommand\seconddot{\rlap.\second}
\newcommand{\myurl}[1]{{\tt #1}}

\newcommand\pho{\phantom0}
\newcommand\plm{$\pm$}

\newcommand{\massasym}[3]{#1\rlap{$^{+#2}_{-#3}$}{\phantom{\plm0.00}}}
\newcommand{\tempsym}[2]{#1\rlap{\plm#2}{\phantom{\plm0.0}}}
\newcommand{\tempasym}[3]{#1\rlap{$^{+#2}_{-#3}$}{\phantom{\plm0.0}}}

\newcommand\qpah{\ensuremath{q_{\mbox{\scriptsize\sc pah}}}}
\newcommand\Umin{\ensuremath{U_{\rm min}}}
\newcommand\Umax{\ensuremath{U_{\rm max}}}
\newcommand\Mdust{\ensuremath{M_{\rm d}}}
\newcommand\Ldust{\ensuremath{L_{\rm d}}}
\newcommand\Tdust{\ensuremath{T_{\rm d}}}

\usepackage{textcomp}
\renewcommand{\micron}{\hbox{\textmu}m}
\let\um=\micron

\begin{document}

\title{BLAST Observations of Resolved Galaxies: Temperature Profiles and the
Effect of AGN on FIR to Submillimeter Emission}

\author{
Donald~V.~Wiebe,\altaffilmark{1,2}
Peter~A.~R.~Ade,\altaffilmark{3}
James~J.~Bock,\altaffilmark{4,5}
Edward~L.~Chapin,\altaffilmark{1}
Mark~J.~Devlin,\altaffilmark{6}
Simon~Dicker,\altaffilmark{6}
Matthew~Griffin,\altaffilmark{3}
Joshua~O.~Gundersen,\altaffilmark{7}
Mark~Halpern,\altaffilmark{1}
Peter~C.~Hargrave,\altaffilmark{3}
David~H.~Hughes,\altaffilmark{8}
Jeff~Klein,\altaffilmark{6}
Gaelen~Marsden,\altaffilmark{1}
Peter~G.~Martin,\altaffilmark{9,10}
Philip~Mauskopf,\altaffilmark{3}
Calvin~B.~Netterfield,\altaffilmark{2,10}
Luca~Olmi,\altaffilmark{11,12}
Enzo~Pascale,\altaffilmark{3}
Guillaume~Patanchon,\altaffilmark{13}
Marie~Rex,\altaffilmark{6}
Douglas~Scott,\altaffilmark{1}
Christopher~Semisch,\altaffilmark{6}
Nicholas~Thomas,\altaffilmark{7}
Matthew~D.~P.~Truch,\altaffilmark{6}
Carole~Tucker,\altaffilmark{3}
Gregory~S.~Tucker,\altaffilmark{14}
Marco~P.~Viero\altaffilmark{10}
}

\altaffiltext{1}{Department of Physics \& Astronomy, University of British Columbia, 6224 Agricultural Road, Vancouver, BC V6T~1Z1, Canada}
\altaffiltext{2}{Department of Physics, University of Toronto, 60 St. George Street, Toronto, ON M5S~1A7, Canada}
\altaffiltext{3}{School of Physics and Astronomy, Cardiff University, 5 The Parade, Cardiff, CF24~3AA, UK}
\altaffiltext{4}{Jet Propulsion Laboratory, Pasadena, CA 91109-8099}
\altaffiltext{5}{Observational Cosmology, MS 59-33, California Institute of Technology, Pasadena, CA 91125}
\altaffiltext{6}{Department of Physics and Astronomy, University of Pennsylvania, 209 South 33rd Street, Philadelphia, PA 19104}
\altaffiltext{7}{Department of Physics, University of Miami, 1320 Campo Sano Drive, Coral Gables, FL 33146}
\altaffiltext{8}{Instituto Nacional de Astrof\'{\i}sica \'Optica y Electr\'onica, Aptdo. Postal 51 y 216, 72000 Puebla, Mexico}
\altaffiltext{9}{Canadian Institute for Theoretical Astrophysics, University of Toronto, 60 St. George Street, Toronto, ON M5S~3H8, Canada}
\altaffiltext{10}{Department of Astronomy \& Astrophysics, University of Toronto, 50 St. George Street, Toronto, ON  M5S~3H4, Canada}
\altaffiltext{11}{Istituto di Radioastronomia, Largo E. Fermi 5, I-50125, Firenze, Italy}
\altaffiltext{12}{University of Puerto Rico, Rio Piedras Campus, Physics Dept., Box 23343, UPR station, San Juan, Puerto Rico}
\altaffiltext{13}{Laboratoire APC, 10, rue Alice Domon et L{\'e}onie Duquet 75205 Paris, France}
\altaffiltext{14}{Department of Physics, Brown University, 182 Hope Street, Providence, RI 02912;}

\begin{abstract}
Over the course of two flights, the Balloon-borne Large Aperture Submillimeter
Telescope (BLAST) made resolved maps of seven nearby ($<25$\,Mpc) galaxies
at 250, 350, and 500\,\um.
During its June 2005 flight from Sweden, BLAST observed a single
nearby galaxy, NGC~4565.   During the December 2006 flight from Antarctica,
BLAST observed the nearby galaxies NGC~1097, NGC~1291, NGC~1365,
NGC~1512, NGC~1566, and NGC~1808.  We fit physical dust models to a combination
of BLAST observations and other available data for the galaxies observed by
\Spitzer.
We fit a modified blackbody to the remaining galaxies
to obtain total dust mass and mean dust temperature.  For the four galaxies with
\Spitzer\ data, we also produce maps and radial profiles of dust column density
and temperature.  We measure the fraction of BLAST detected flux originating
from the central cores of these galaxies and use this to calculate a ``core
fraction,'' an upper limit on the ``AGN fraction'' of these galaxies.
We also find our resolved observations of these galaxies give a dust mass
estimate 5--19 times larger than an unresolved observations would predict.
Finally, we are able to use these data to derive a value for the
dust mass absorption co-efficient of
$\kappa=0.29\pm0.03$\,m$^2$\,kg$^{-1}$ at 250\,\micron.  This study is
an introduction to future higher-resolution and
higher-sensitivity studies to be conducted by \Herschel\ and \SCUBATOO.
\end{abstract}


\section{Introduction}
\label{sec:intro}

\begin{figure*}
\centering
\includegraphics[width=6in]{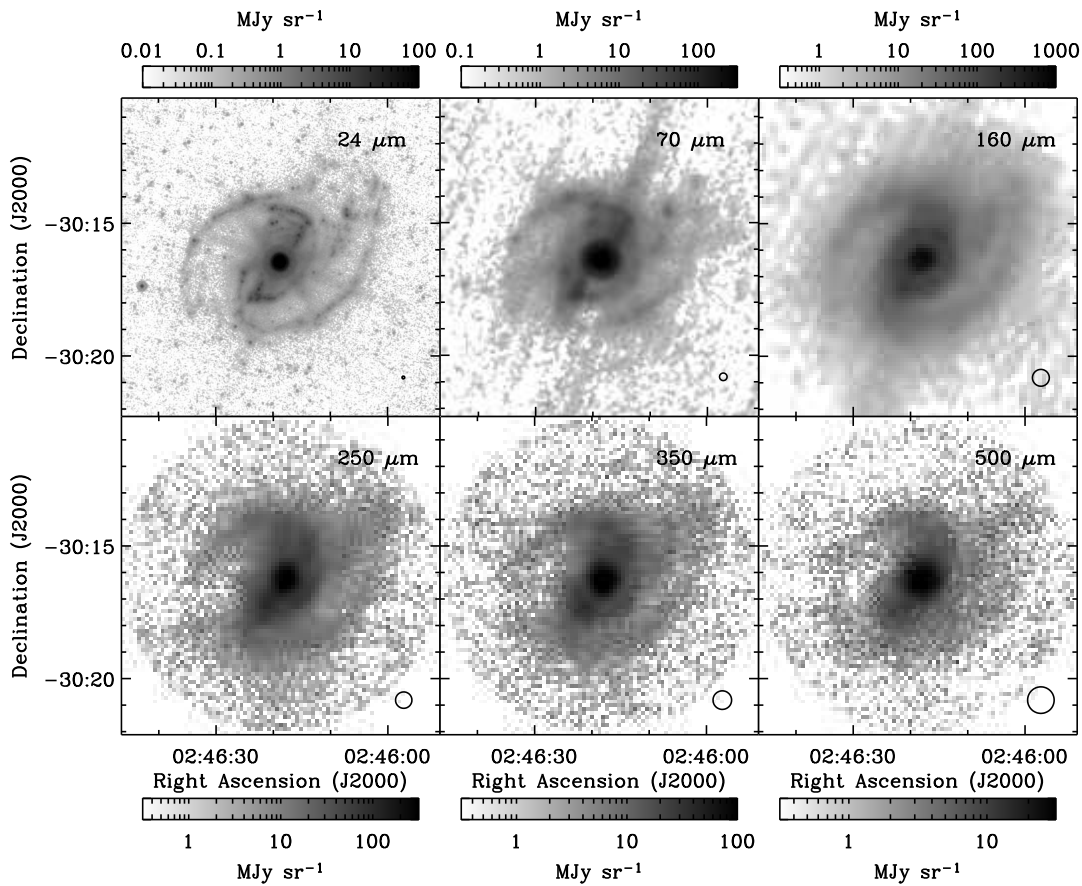}
\caption{%
%BLAST observations of NGC~1097.  Pixels are 9\arcsec\ on a side.  The radius of
%the clipped map is 6\arcmin.  Note the logarithmic intensity scale.  The size
%of the BLAST and MIPS beams are plotted for reference in the lower right corner.
BLAST ({\it bottom row\/}) and MIPS ({\it top row\/}) observations of NGC~1097. 
The radius of the clipped BLAST map is
6\arcmin.  The intensity scale is logarithmic.
The FWHM size
of the BLAST and MIPS beams are plotted for reference in the lower right corner.
BLAST resolves the spiral arms and central bar of the galaxy.  The 24\,\um\
image is shown for comparison only, and not used in this analysis.
\label{fig:ngc1097}
}
\end{figure*}

Much of the bolometric luminosity produced by entire galaxies is
emitted as thermal radiation from dust grains in the interstellar
medium (ISM) at temperatures of several tens of Kelvin. The heat
source is predominantly optical and ultra-violet (UV) light from
stars, as well as matter accretion onto an active galactic nucleus
(AGN) in some objects. In either case, the spectrum of this light
tends to peak in the range $\sim$60--200\,\micron. Since the launch of
the {\it Infrared Astronomical Satellite\/} (\IRAS) in 1983,
observations at these wavelengths have been essential in a number of
fields, ranging from the studies of the earliest stages of
star-formation in dusty molecular clouds in our own Galaxy, to the
detection of distant star-forming galaxies that emit most of their
light in this previously unexploited band. Furthermore, excluding
light from the Cosmic Microwave Background, it is known that
approximately half of all of the light in the Universe is emitted at
these wavelengths, the Cosmic Infrared Background (CIB) \citep[see
also \citealt{devlin2009}]{puget1996, fixsen1998}.

Observations from the ground are impossible at wavelengths in
the range
$\sim$60--200\,\micron,
the far infrared (FIR) band,
due primarily to the
absorption and emission from atmospheric water vapor. Therefore, most
of the progress at these wavelengths since {\it IRAS\/} has also been
undertaken by satellites: the {\em Infrared Space Observatory}
(\ISO), the {\em Spitzer Space Telescope\/}, and the {\em Akari
Space Telescope}.  However, the lack of data at longer wavelengths
can bias the interpretation of these results toward the assumption of
warmer dust temperatures, resulting in underestimates of luminosity and
mass.  At slightly longer wavelengths, $\sim$350--1000\,\micron, the
submillimeter band, several spectral windows are
available from dry, high-altitude sites such as Mauna Kea or the
Atacama Desert. In particular, studies at 850\,\micron\ with SCUBA
\citep[e.g.][]{smail1997, hughes1998, holland1999, eales2000, dunne2001,
scott2002, mortier2005, stevens2005, nutter2007} have been used to detect very
cool dust, both in our Galaxy, and in more distant galaxies.
%However, typically 
%data at only a single wavelength are available, and a dust temperature
%must be assumed
%in order to estimate temperatures and luminosities.
Typically in these surveys a dust temperature must be assumed in order to
estimate redshifts and luminosities, because data are available only at a
single wavelength.  Therefore the
combination of ground-based submillimeter observations with space-based FIR
measurements has been required to assess the bolometric luminosities
and apparent dust temperatures of everything from individual
pre-stellar objects, up to entire galaxies at redshifts
$z>1$. However, the general lack of high signal-to-noise data in the crucial
wavelength range
$\sim$200--500\,\micron\ results in large uncertainties in these quantities.

In this paper we present results from a study at 250, 350, and
500\,\micron\ of several nearby galaxies using the 2-m Balloon-borne
Large Aperture Submillimeter Telescope (BLAST). BLAST was built to
fill the wavelength gap between ground-based submillimeter and space-based
FIR instrumentation, using a prototype of the SPIRE camera for
\Herschel\ \citep{griffin2007}.

%During its first scientific flight, hereafter BLAST05,
%\citep{pascale2008},
In June 2005, BLAST was launched
from Esrange, Sweden, for it's first scientific flight, hereafter BLAST05.
It acquired data for 100 hours before landing on
Victoria Island, Canada.  %For this flight BLAST was equipped with a
%2-m spherical primary mirror.  The flight was characterized by
%point-source sensitivity significantly below expectations
%\citep{truch2008}.  As a result, observations during BLAST05
%concentrated primarily on Galactic surveys \citep[{\it
%e.g.\/},][]{chapin2008}.
During this flight, BLAST observed the nearby galaxy NGC~4565.
%\citep{dreyer1888}
%
In December 2006, BLAST conducted its second science flight, BLAST06,
carrying out 250 hours of observations above Antarctica.
%  Several
%fields were mapped, including a $\sim$50 deg$^2$ Galactic field in the
%direction of Vela \citep{netterfield2009} and a $\sim$10 deg$^2$
%extragalactic field overlapping GOODS-South \citep{devlin2009,
%dye2009, marsden2009, pascale2009}.  Observations of six nearby
%resolved galaxies were also made:
%An additional six nearby galaxies were observed during BLAST's second flight
%(BLAST06):
%\citep[BLAST06, ][]{pascale2008}:
Six nearby galaxies were observed during BLAST06:
NGC~1097, NGC~1291, NGC~1365,
NGC~1512, NGC~1566, and NGC~1808. %\citep{dreyer1888}.

All these galaxies are resolved by BLAST.
The full widths at half-maximum (FWHM) of the BLAST06 beams are 36,
42, and 60\,\arcsec\ at 250, 350, and 500\,\um\ respectively.  The BLAST05
flight suffered from degradation of the warm optics \citep{truch2008}.
The single galaxy in our sample from BLAST05 is nevertheless resolved, despite
the large, non-Gaussian point spread functions due to the degraded optics during
this flight.
We are therefore able to map the temperature and
luminosity distributions,
and hence dust column density, with greater signal-to-noise ratio (SNR) than any
existing work.  Our study can be considered a precursor to The
\Herschel\ Reference Survey \citep{boselli2008} and the JCMT Nearby
Galaxies Legacy Survey \citep{wilson2009} that will be undertaken by
\Herschel\ and \SCUBATOO, respectively.  A complete description of the BLAST
instrument is given in \citet{pascale2008}.

The known correlation between SED shape and luminosity in \IRAS\ galaxies
\citep[e.g.][]{soifer1991} means good measurements of these quantities for
rest-frame galaxies are required to put proper constraints on high-redshift
FIR source count models (e.g.~Chapin et al., in prep.).
%\citep[e.g.][]{chapin2009}.
BLAST samples the Rayleigh-Jeans side of
thermal emission from local galaxies; when combined with FIR measurements
from other instruments, the dust thermal emission peak is bracketed, allowing
robust determination of these quantities.
%{\bf Known correlation between temperature and luminosity important to local
%luminosity function (see intro to Chapin, Hughes and Aretxaga 2009 for
%references) -- need to measure luminosity and temperature accurately for local
%sample before we try to evolve it to model the high-z population.}

\begin{figure*}
\centering
\includegraphics[width=6in]{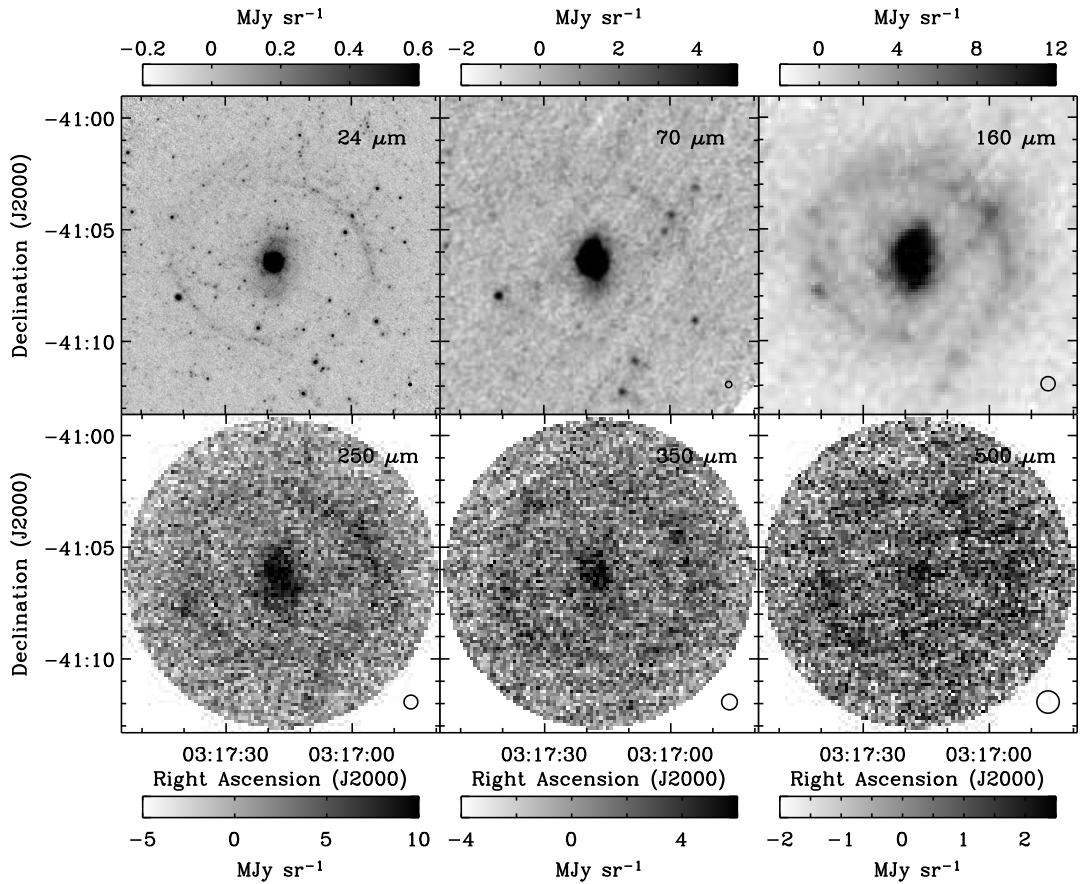}
\caption{%
BLAST ({\it bottom row\/}) and MIPS ({\it top row\/}) observations of NGC~1291. 
The radius of the clipped map BLAST map is 7\arcmin.
The intensity scale is linear.  The size
of the BLAST and MIPS beams are plotted for reference in the lower right corner.
The outer 4\arcmin\ ring of spiral arm structure is
detected in all three BLAST bands, as is the central core.
%BLAST observations of NGC~1291.  The pixel size is 9\arcsec\ on a side.  The
%radius of the clipped map is 7\arcmin.  The outer ring is detected in all three
%bands.  Note the linear intensity scale.
\label{fig:ngc1291}
}
\end{figure*}

Resolved studies of galaxies are required to better model galactic structure
and dynamics.  The unprecedented ability of BLAST to make high resolution
maps on large angular scales in the submillimeter band provides the best
opportunity to-date to study the structure of diffuse dust in nearby galaxies.
To properly estimate star formation rates from submillimeter fluxes of
unresolved, high redshift galaxies, an understanding of the fraction of
submillimeter flux originating from AGN-driven dust heating is required.
Although BLAST is unable to distinguish AGN-driven dust heating from nuclear
starburst driven heating, and this study involves nearby spirals, rather
than high-redshift starbursts, we are still able to put limits on this AGN
fraction by comparing flux measured from the core of these resolved galaxies
with their total integrated flux.
%{\bf Resolved studies of galaxies required for galaxy models:
%where are the stars forming, how much light is produced in each
%region etc. -- BLAST does the best job of this to date.}
Furthermore, because BLAST can make high-fidelity large-area maps, it is able to
measure the complete emission from these spatially extended galaxies.
%This gives
%BLAST access to galaxies too large to fully sample with SCUBA-2.
%{\bf BLAST can integrate over large areas and get the fluxes right.}
%{\bf SLUGS SCUBA local luminosity function needs to reject the closest
%objects because can't handle resolved structure.}

%The remainder of this paper is organized as follows: Section~\ref{sec:dr}
%reviews the reduction of the BLAST data; Section~\ref{sec:obs} describes the
%data used in this study, from BLAST and other experiments; Section~\ref{sec:sed}
%explains our method of fitting models to the data; 

\section{Data Reduction}
\label{sec:dr}
%A description of the BLAST experiment has been presented in \citet{pascale2008},
The common
data reduction pipeline for BLAST has been presented in \citet{truch2008}
and \citet{patanchon2008}.  Maps used in this analysis for NGC~4565 were made
using the maximum-likelihood map maker \SANEPIC\ \citep{patanchon2008} which was
written explicitly for BLAST analysis.

Maps of BLAST observations of the remaining galaxies were produced using
\Almagest\ \citep{wiebe2008}, an iterative, maximum-likelihood map maker
also written for BLAST
analysis, and algorithmically similar to other iterative map makers used in CMB
analysis
(e.g.~\citealt{prunet2000}).  Because most of these galaxies were observed
only at a single scan direction, and hence lack cross-linking, the maps
produced
are artificially constrained to zero flux far from the observed galaxy to
improve convergence properties of the iterative algorithm (see \S 3.8 of
\citealt{patanchon2008}).  In relatively high SNR cases, such as
these galaxy maps, \Almagest\ output is consistent with maps made with
\SANEPIC\ \citep{wiebe2008}, but may be produced faster, and allow arbitrary
constraints to be applied simply.

\begin{figure*}
\centering
\includegraphics[width=6in]{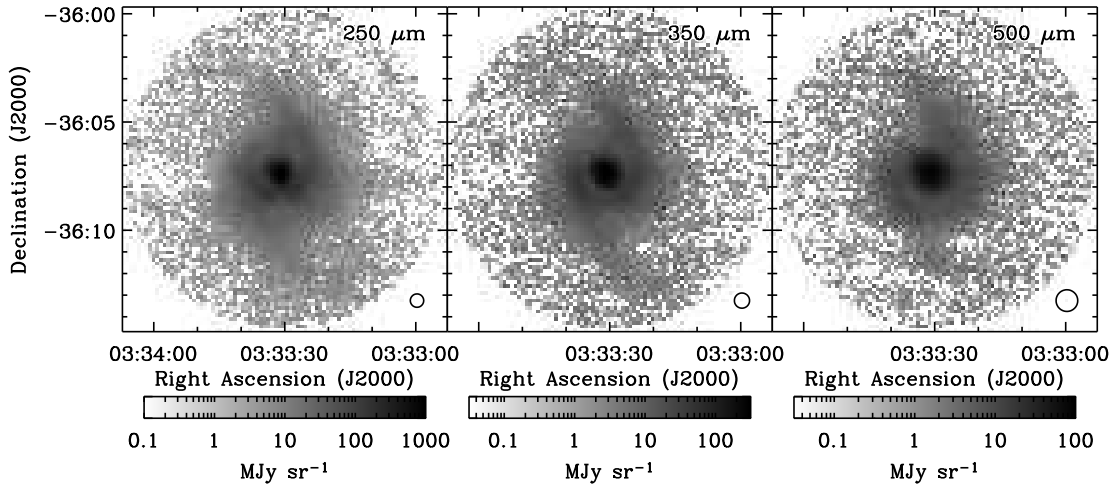}
\caption{
BLAST observations of NGC~1365.  Pixels are 9\arcsec\ on a side.  The radius of
the clipped map is 6\farcm5.  The intensity scale is logarithmic.
%{\it Top row:\/} BLAST observations of NGC~1365.  The radius of the clipped map
%is 6\farcm5.
%{\it Bottom row:\/} BLAST observations of NGC~1808.
%In the map of NGC~1808, BLAST also
%detects excess flux near the position of the interacting galaxies ESO 305-IG 010
%($\alpha=5^{\rm h}08^{\rm m}27^{\rm s}$, $\delta=-37^\circ39\farcm5$; lower
%left).
%For both galaxies, the intensity scale is logarithmic.
\label{fig:ngc1365}
}
\end{figure*}

\section{Observations}
\label{sec:obs}

\subsection{BLAST Observations}
The seven nearby galaxies observed by BLAST were selected based on a number of
competing criteria.  First,
%they had to be visible to BLAST during its flight
%(i.e., in the northern hemisphere during BLAST05 and in the southern
%hemisphere during BLAST06).
the visibility requirements were restrictive,
as they took into account BLAST's limited elevation range (25\dg--60\dg), Sun
and Moon avoidance criteria, and still had to make allowance for the changing
latitude and longitude of the telescope arising from BLAST's uncertain ground
path while aloft.  Second, the galaxies had to be sufficiently large to be
resolved by BLAST's optics (e.g. BLAST's 36\arcsec\ beam size at 250\,\um\ 
corresponds to $\sim$2.5~kpc at the average distance of 14~Mpc for this galaxy
sample).  Third, they had to be sufficiently luminous for
BLAST to be able to map them with a high SNR in the relatively small
fraction of the flights scheduled for these observations.  Finally, the sample
needed
to have measurements from the {\it Infrared Astronomical Satellite} (\IRAS), the
{\it Spitzer Space Telescope}, the Submillimeter Common-User Bolometer Array
(SCUBA), or other experiments complementary to BLAST.

Table~\ref{tab:physparm} summarizes relevant physical parameters of the BLAST
nearby galaxy sample\rlap.\footnote{Instructions for
obtaining BLAST maps used in this analysis may be found on
\url{http://blastexperiment.info/}.}
The measured total flux density in each
BLAST band is tabulated in Table~\ref{tab:fluxen}.

\subsection{Other Multiwavelength Data}
\label{sec:multi}

As part of the \Spitzer\ Infrared Nearby Galaxies Survey
\citep[SINGS;][]{kennicutt2003},
four of our sample (NGC~1097, NGC~1291, NGC~1512, and NGC~1566, hereafter
BLAST/SINGS galaxies), have also been
observed by the Multiband Imaging Photometer for \Spitzer\ (MIPS) at
160\,\um, 70\,\um, and 24\,\um\
\citep{rieke2004} and the Infrared Array Camera (IRAC)
at 8\,\um, 6\,\um, 4.5\,\um, and
3.6\,\um\ \citep{fazio2004}, both on the {\it Spitzer Space Telescope}.  The
remaining
galaxies in our sample were not observed by \Spitzer.  Spatially integrated
\Spitzer\ fluxes
are taken from \citet{dale2007}, and MIPS maps used in
this analysis come from the fifth SINGS public data
release\rlap.\footnote{obtained
from \myurl{http://data.spitzer.caltech.edu/popular/ sings/}}

\IRAS\ data for all of our sample except NGC~1291 have been extracted from
SCANPI\rlap.\footnote{available at
\myurl{http://scanpi.ipac.caltech.edu:9000/ applications/Scanpi/index.html}}
\IRAS\ data for the fainter and more extended NGC~1291 are measured from HIRES
generated maps \citep{aumann1990}.
These \IRAS\ data are used, in addition to \Spitzer, to constrain the spatially
integrated SEDs of these galaxies.
A 20\% uncorrelated uncertainty is applied to all \IRAS\ data.
\IRAS\ 25\,\um\ data
are not used in this analysis when equivalent MIPS 24\,\um\ data are available.

Two of our sample have been observed by SCUBA: NGC~1097 at 850\,\um\
\citep{dale2005},
and NGC~1808 at both 450 and 850\,\um\ \citep{stevens2005}.  These measurements
do not cover the entire region of emission detected by BLAST and we omit these
measurements from our fits for this reason.

For NGC~1808, we calculate
aperture corrections for the \citet{stevens2005} reported flux densities.
We apply the \citeauthor{stevens2005} apertures to the BLAST observations and
compare the
500\,\um\ flux inside these apertures to the total 500\,\um\ flux.  From this,
we calculate flux densities for NGC~1808 of $s_{450}=13.12\,\rm Jy$ and
$s_{850}=1.76\,\rm Jy$ by applying the aperture corrections to the SCUBA
measurements.
%of 1.84 and 1.35 at 450\,\um\ and 850\,\um, respectively.
These values are
consistent with our $\beta=2$ modified blackbody model fit to BLAST plus
100\,\um\ IRAS (\S\,\ref{sec:globsed}).  The aperture-corrected SCUBA
measurements are plotted in Fig.~\ref{fig:glosed1}, along with the model fit.

We also exclude data from the {\it
Infrared Space Observatory\/} (\ISO) for NGC~1365 and NGC~4565, again because
of aperture effects.

\subsection{Individual Galaxies}
A brief description of each observed galaxy follows.
\subsubsection{NGC~1097}
Features of NGC~1097, presented in Fig.~\ref{fig:ngc1097},
include a bar 20\,kpc in length that tapers off into two spiral arms
\citep{beck2005}, as well as several fringe structures called jets or `rays'
\citep{wolstencroft1975,lorre1978}.
%NGC~1097 has a dwarf galaxy companion,
%NGC~1097A, which is at a projected distance of 18\,kpc to its north-west and is
%morphologically an elliptical E5 \citep{ondrechen1989,beck2005,
%wolstencroft1975,sersic1965}.
NGC~1097 has an active Seyfert~1/LINER nucleus,
%that emits broad double-peaked balmer emission lines that
%are evidence for an accretion disk around a central black hole
%(?)\citep{storchi-bergmann1993,storchi-bergmann2003,nemmen2004}.  
which is considered to have a low luminosity \citep{mason2007}, and is
surrounded by a 1.5\,kpc diameter nuclear ring, containing
$1.1\times10^9$\,M$_\odot$ of molecular gas \citep{hsieh2008,gerin1988}.  
Thermal dust heating in the nuclear ring is dominated by a starburst
\citep{mason2007} with a stellar production rate of 5\,M$_\odot$\,yr$^{-1}$ 
\citep{hummel1987}.  BLAST made two observations of a \hbox{$\sim$0.25}\,\sqdg\
area centered on NGC~1097, with a total integration time of 52~minutes.
BLAST clearly resolves the bar and spiral arms of the galaxy, but has
insufficient resolution to resolve the 1.5~kpc star forming ring.

\begin{figure*}
\centering
\includegraphics[width=6in]{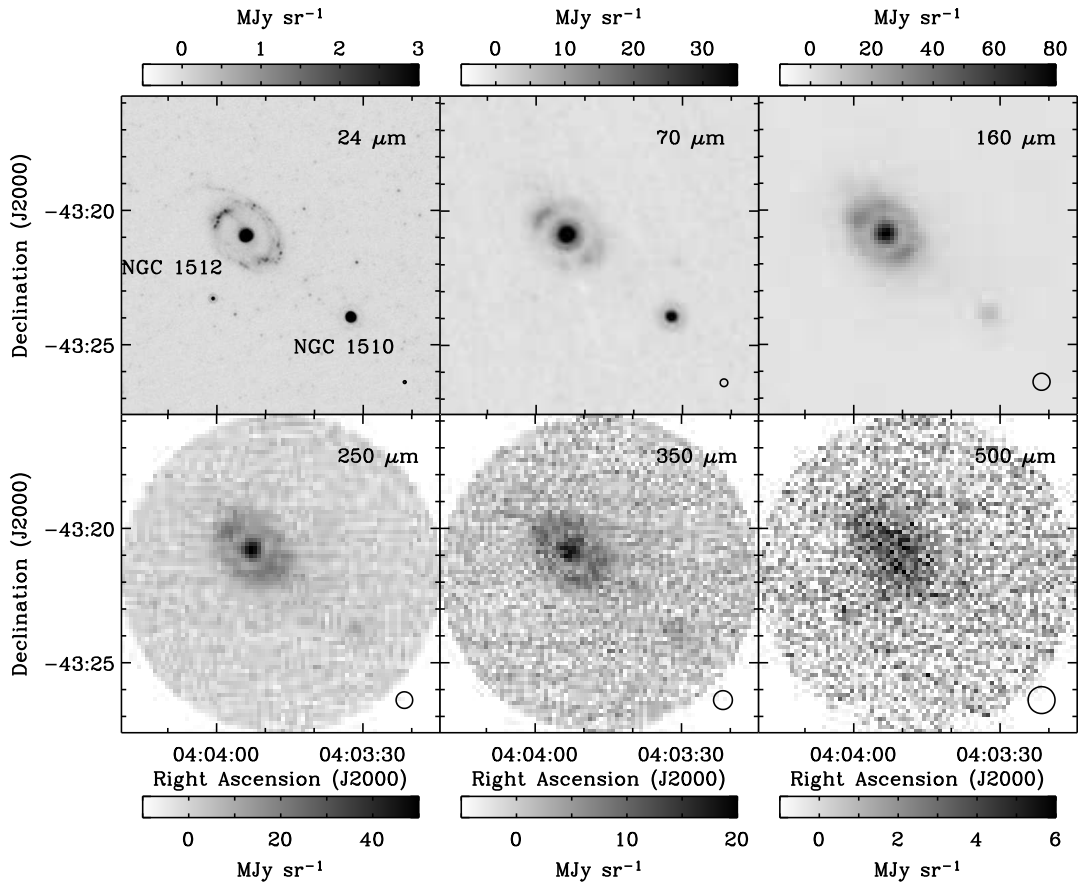}
\caption{
%BLAST observations of NGC~1512 (upper left).  The pixel size is 9\arcsec\ on a
%side.  The radius of the clipped map is 6\arcmin.  Also detected
%by BLAST at 250\micron\ and 350\micron\ is NGC~1510
%($\alpha=4^{\rm h}03^{\rm m}33^{\rm s}$, $\delta=-43^\circ24^\prime$; lower
%right).  Note the linear intensity scale.
BLAST ({\it bottom row\/}) and MIPS ({\it top row\/}) observations of NGC~1512. 
The radius of the clipped BLAST maps is 6\arcmin.  
The intensity scale is linear.  The size
of the BLAST and MIPS beams are plotted for reference in the lower right corner.
BLAST detects the companion galaxy, NGC~1510
($\alpha=4^{\rm h}03^{\rm m}33^{\rm s}$, $\delta=-43^\circ24^\prime$; lower
right)
at 250\,\micron\ and 350\,\micron.  In these two bands, BLAST also can
distinguish between the core and disk of NGC~1512.
\label{fig:ngc1512}
}
\end{figure*}

\subsubsection{NGC~1291}
NGC~1291, presented in Fig.~\ref{fig:ngc1291},
%is classified as a R$^\prime$SB(s)0/a galaxy by
%\citet{devaucouleurs1975}, but
%several other authors have given it slightly different classifications
%\citep[{e.g.}][]{vandriel1988}.  This
is the
brightest Sa galaxy in the Shapley-Ames Catalogue \citep{bregman1995}.
NGC~1291 has an
orientation that is almost `face on' and has a primary bar and secondary bar
that have a 30\dg\ misalignment with each other \citep{perez2006}.
%The outer
%ring
%of the galaxy has two spiral arms that are faintly apparent \citep{vandriel1988}
The total \ion{H}{1} mass is
$2.4\times10^9$\,M$_\odot$, and there is little evidence for star formation in
the central bulge \citep{hogg2001}.  BLAST made four observations of a
\hbox{$\sim$0.4}\,\sqdg\ area centered on NGC~1291, with a total integration
time of 92~minutes.  Although diffuse, BLAST detects both the central core of
the galaxy, as well as the ring-like spiral structure 4\arcmin\ from the core.

\begin{figure*}
\centering
\includegraphics[width=6in]{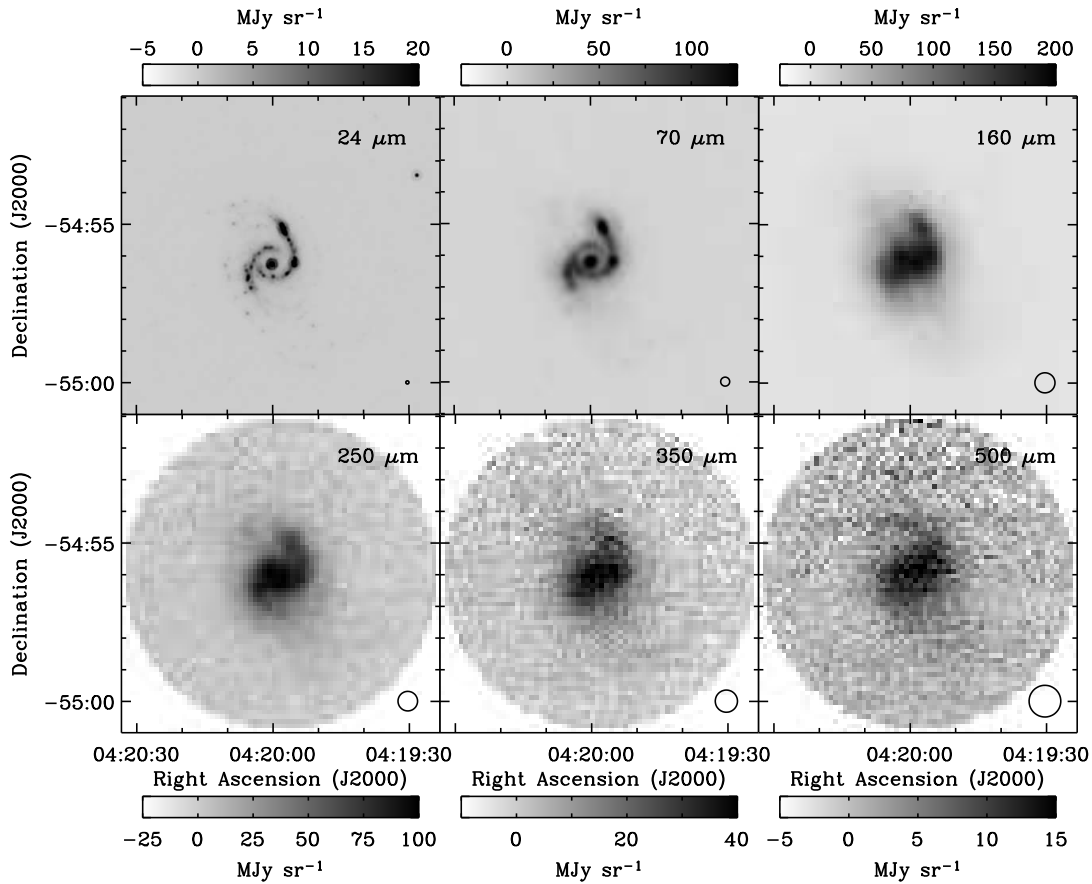}
\caption{
%BLAST observations of NGC~1566.  Pixels are 9\arcsec\ on a side.  The radius of
%the clipped map is 5\arcmin.
%Note the linear intensity scale.
BLAST ({\it bottom row\/}) and MIPS ({\it top row\/}) observations of NGC~1566. 
The radius of the clipped BLAST maps is 5\arcmin.  
The intensity scale is linear.  The size
of the BLAST and MIPS beams are plotted for reference in the lower right corner.
BLAST detects the spiral arms, although the map is dominated by flux from the
core.
\label{fig:ngc1566}
}
\end{figure*}

\subsubsection{NGC~1365}
NGC~1365, presented in Fig.~\ref{fig:ngc1365}, is
%a spiral galaxy classified by Sandage and Tammann as SBb(s)I.
%Both broad and narrow Balmer emission lines are present in radiation coming
%from the active nucleus, along with high excitation line emission.  Because of
%this
is usually classified as a Seyfert 1.5, but some authors list
it as Seyfert~1 or Seyfert~2
\citep[e.g.,][]{galliano2005,veron1980,alloin1981}.  The
AGN has low luminosity, but there is evidence from optical studies to suggest
that there is star formation surrounding the nucleus \citep{komossa1998}.  BLAST
made three observations of a \hbox{$\sim$0.25}\,\sqdg\ area centered
on NGC~1365, with a total integration time of 30~minutes.  The brightest of
our sample, BLAST's observations prominently show both the bar and spiral arm
structure of the galaxy.

\subsubsection{NGC~1512}
NGC~1512, presented in Fig.~\ref{fig:ngc1512},
%is classified as SB(r)a \citep{devaucouleurs1976}.  Its nucleus is
has a nucleus
surrounded by a circumnuclear ring \citep{maoz2000}.  Beyond the ring, its
spiral arms are complicated by its satellite galaxy, NGC~1510
\citep{kinman1978}.
BLAST made three observations of a \hbox{$\sim$0.45} deg$^2$ area centered
on NGC~1512, with a total integration time of 28~minutes.  BLAST detects the
core and tightly wound central spiral of NGC~1512.  BLAST also detects
the companion galaxy, NGC~1510.

\subsubsection{NGC~1566}
NGC~1566, presented in Fig.~\ref{fig:ngc1566},
%is an SAB(s)bc galaxy with an active Seyfert 1 nucleus.
%Evidence of UV
%thermal heating to \ion{H}{2}, but Reunanen suggest that only 10\% of UV
%excitation is
%from starburst and most of it is from \ion{H}{2} gas collision from shock
%\citep{reunanen2002}.
is the second brightest known Seyfert galaxy \citep{devaucouleurs1973}.
Extending from the nucleus are two prominent spiral arms which
continue outward to form the galaxy's outer ring that is at an inclination of
about
30\degr\ \citep{devaucouleurs1975,bottema1992}.  The arms contain a significant
amount of star formation \citep{bottema1992}.
BLAST made two observations of a \hbox{$\sim$0.7} deg$^2$ area centered on
NGC~1566, with a total integration time of 50~minutes.  BLAST resolves
the spiral structure of the galaxy, and the observations are dominated by
submillimeter flux originating in the core of the galaxy.

\subsubsection{NGC~1808}
\begin{figure*}
\centering
\includegraphics[width=6in]{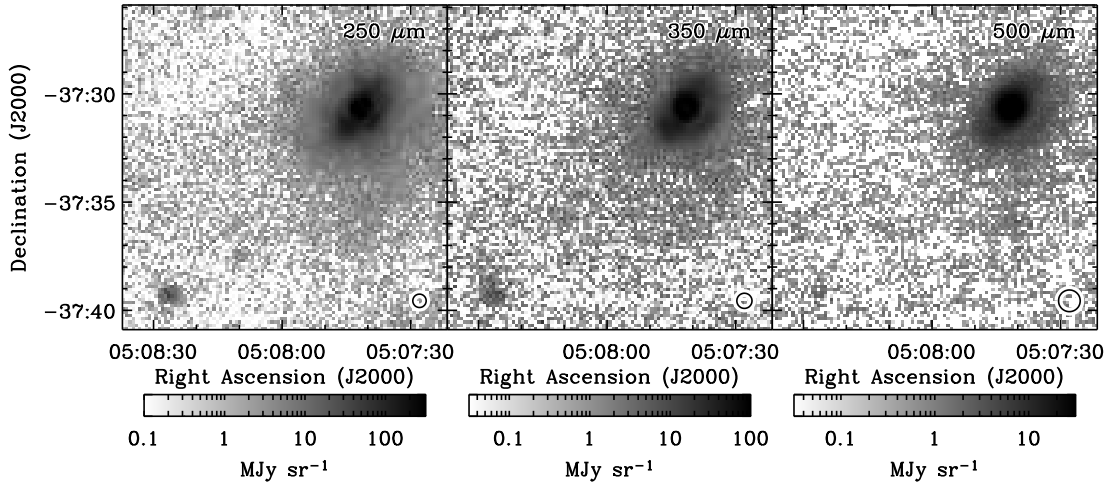}
\caption{
BLAST observations of NGC~1808 (upper right).  Pixels are 9\arcsec\ on a side.
BLAST also
detects excess flux density near the position of the interacting galaxies
ESO 305-IG~010
($\alpha=5^{\rm h}08^{\rm m}27^{\rm s}$, $\delta=-37^\circ39\farcm5$,
$z=0.0524$; lower left).  The intensity scale is logarithmic.
\label{fig:ngc1808}
}
\end{figure*}

NGC~1808, presented in Fig.~\ref{fig:ngc1808},
is a
%classified as an SABb
Seyfert 2 galaxy \citep{jimenez-bailon2005}.
%Observation of optical
%emission lines coming from the nucleus indicate that this is a Seyfret 2
%galaxy.r
Subsequent measurement
%done in both IR (\ISO) and X-ray (XMN-Newton,
%\Chandra, \ROSAT) part of the electromagnet spectrum also
seems to indicate that a
large part of the radiation emanating from the central 1\,kpc of the galaxy is
from an active galactic nucleus as well as a high rate of star formation
\citep{foerster_schreiber2003,maiolino2003}.
%Outside of the nucleus there are several luminous hot spot in the radio and IR
%part which appear to be super nova remnants
%\citep[, get first reference]{jimenez-bailon2005}
Using \ROSAT,
\citet{junkes1995} estimate a star formation rate of between 5 and
13\,M$_\odot$\,yr$^{-1}$, and a supernova event rate of 0.09\,yr$^{-1}$.
A companion
galaxy, not observed by BLAST, NGC~1792 is located 40\arcmin\ from NGC~1808
(130\,kpc
at a distance of 10.9\,Mpc) and may be responsible for an accelerated star
formation rate in NGC~1808 \citep{jimenez-bailon2005}.
BLAST made four observations of a \hbox{$\sim$1}~deg$^2$ area centered on
NGC~1808, with a total integration time of 179~minutes.  BLAST detects both the
core and disk of the galaxy.  BLAST also detects excess flux in this map near
the location of the interacting galaxies 
ESO 305-IG~010 ($\alpha=5^{\rm h}08^{\rm m}27^{\rm s}$,
$\delta=-37^\circ39\farcm5$, $z=0.0524$).

\subsubsection{NGC~4565}
\begin{figure*}
\centering
\includegraphics[width=6in]{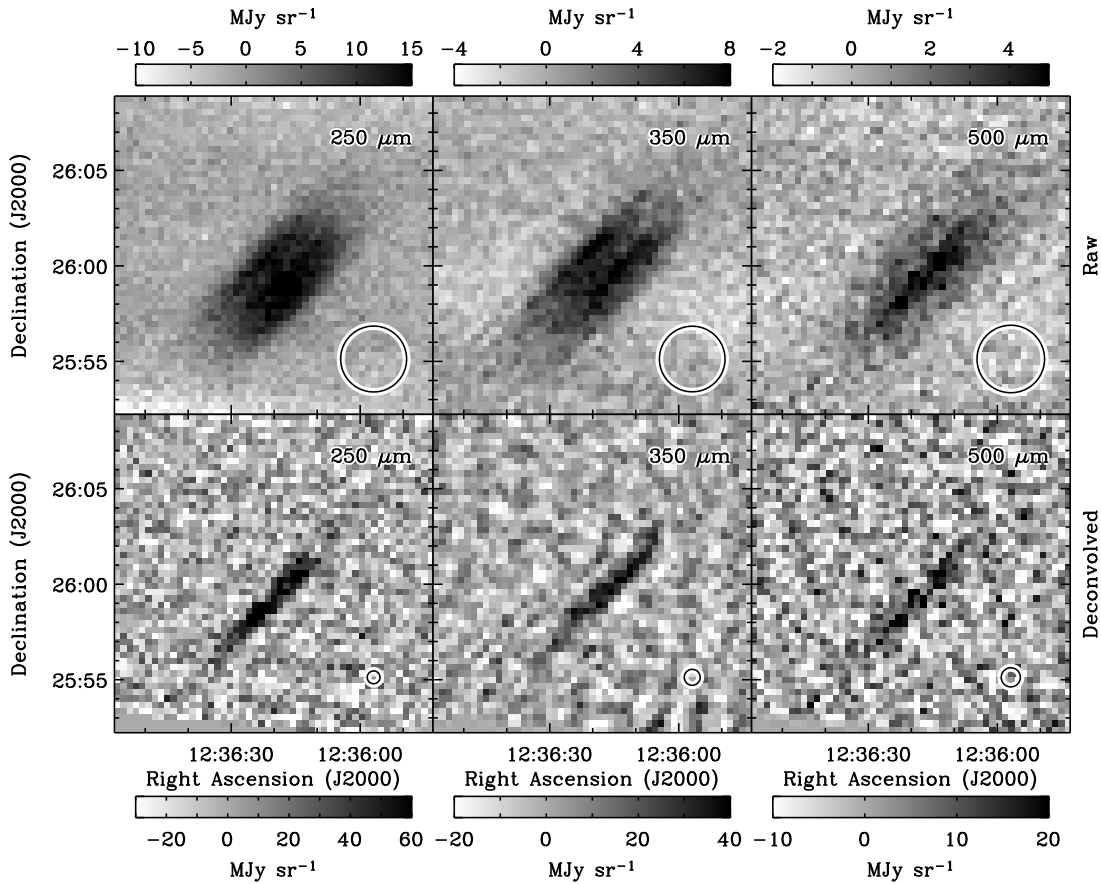}
\caption{
{\it Top row:\/} BLAST observations of NGC~4565.
{\it Bottom row:\/}
Results of deconvolving the raw maps following the procedure outlined in \S2.4
of \citet{chapin2008}.  This deconvolution process results in
3--5 times better resolution, at a cost of reduction in signal-to-noise of a
factor of $\sim$2.
Pixels are 20\arcsec\ on a side.
Note the linear intensity scale.
Effective beam sizes, based on the full widths at half-power (FWHP) found in
Table~1 of \citet{chapin2008}, are plotted in the lower right corners for both
the raw and deconvolved maps.
The full BLAST05 point spread functions are shown
in Figure~1 of \citet{truch2008}.
\label{fig:ngc4565}
}
\end{figure*}

NGC~4565, presented in Fig.~\ref{fig:ngc4565},
is an edge-on barred-spiral galaxy in the
constellation Coma Berenices.  It has an apparent
size of about 12\arcmin\ $\times$ 2\farcm5.
Although degraded by the large point-spread functions of BLAST05
\citep[186--189\arcsec, ][]{truch2008}, NGC~4565 is still resolved by BLAST.\@
BLAST made one observation of a \hbox{$\sim$0.4}~deg$^2$ area centered on
NGC~4565, with a total integration time of 49~minutes.
A beam-deconvolved image, using the method described in \S 2.4 of
\citet{chapin2008}, is shown in the bottom row of Fig.~\ref{fig:ngc4565}.

We fit an exponential profile to the beam-deconvolved maps of
the form $\exp(|x| / s)$, where $x$ is a spatial variable aligned along the
major axis of the galactic disk and $s$ is the scale length, and find
that $s=118$\arcsec, 156\arcsec, and 142\arcsec\ at 250\,\um, 350\,\um,
and 500\,\um, respectively.  The 250\,\um\ fit, which has the highest
signal-to-noise ratio, agrees well with scale lengths measured at other
wavelengths \citep[e.g.,][]{vanderkruit1981,engargiola1992,rice1996}. In
this fit we have ignored the central 3\arcmin\ region of the galaxy.  This
region
corresponds to an area of reduced 250\,\um\ flux density, possibly corresponding
to the molecular ring seen in ${}^{12}$CO \citep{neininger1996}.

\section{SED Fitting}
\label{sec:sed}
\subsection{Physical Dust Models}
\label{sec:dl07}

For the four BLAST/SINGS galaxies, we fit the silicate-graphite-PAH dust models
of \citet{draine2007li}, hereafter {\sl DL07}, to BLAST, MIPS, \IRAS, and IRAC
data.  These models have already been applied, without the benefit of BLAST
observations, to the same galaxies in \citet{draine2007}, hereafter {\sl Dr07}.
We independently recreate this analysis with the inclusion of the BLAST data.
These models provide the dust emissivity per hydrogen atom, $j_\nu(\qpah,
\Umin, \Umax)$, dependent on three model parameters:
\begin{enumerate}
\item \qpah, the fraction of total dust mass in polycyclic aromatic
hydrocarbons (PAHs) containing less than $10^3$ carbon atoms;
\item \Umin, a starlight intensity factor characterizing the radiation field
heating the diffuse ISM;
\item \Umax, a starlight intensity factor characterizing the radiation field
in more intense star forming regions, including photodissociative regions
(PDRs).
\end{enumerate}
Both \Umin\ and \Umax\ are relative to the specific energy density of starlight
measured by \citet{mathis1983}.  Dr07 consider a fourth model
parameter, $\alpha$, which is the exponent characterizing the power-law decrease
in starlight intensity: $d\Mdust/dU\propto U^{-\alpha}$.
They find the model fits to be insensitive to its value, and the models provided
have $\alpha=2$.

Our fitting procedure follows that of Dr07.  We restrict ourselves
to the seven Milky Way dust model sets\footnote{obtained from
\myurl{http://www.astro.princeton.edu/$\sim$draine/ dust/dust.html}}
(\citealt{weingartner2001}, updated by DL07).  These seven model sets have
values of $q_{\rm PAH}$ between $0.47$\% and $4.6$\%.  We take linear
combinations of these to produce 43 different model sets for a finer-grained
sampling of $q_{\rm PAH}$.  Each set contains values of $j_\nu$ for
twenty-two values of \Umin\ between 0.1 and 25, and for five ``PDR models''
with values of \Umax\ ($10^2$, $10^3$, $10^4$, $10^5$, $10^6$), plus one
``diffuse ISM model'' with $\Umax=\Umin$.

These dust models are combined with a stellar blackbody to produce a galactic
model:
\begin{equation}
\begin{split}
F_{\nu,\rm model}=\Omega_\star B_\nu(T_\star) + {\Mdust\over{m_{\rm H} D^2}} 
\left(M_{\rm H}\over\Mdust\right) \\
\times \big[ (1-\gamma) j_\nu(\qpah, \Umin, \Umin) \\
+ \gamma j_\nu(\qpah, \Umin, \Umax) \big]\mbox{,}
\end{split}
\end{equation}
where $\Omega_\star$ is the solid angle subtended by stars, $B_\nu(T)$ is the
Planck function, the nominal stellar temperature $T_\star=5000$\,K, \Mdust\
is the total dust mass, $m_{\rm H}$ the mass of a hydrogen atom, $D$ the
distance to the galaxy, and $\gamma$ a mixing fraction between the diffuse ISM
($\Umax=\Umin$) and PDR ($\Umax>\Umin$) dust models.  The hydrogen-to-dust
mass ratio, $M_{\rm H}/\Mdust$, is set by the model choice, and varies from 96 to
100.

Like Dr07, we find that the best fits for the four galaxies
are insensitive to the choice of \Umax, and so set $\Umax=10^6$.  We obtain
the best-fit values of $\Omega_\star$, \Mdust, $\gamma$, \qpah, and \Umin\
through $\chi^2$ minimization (see \S\,\ref{sec:chi2}).

\begin{figure*}
\centering
\includegraphics[width=6in]{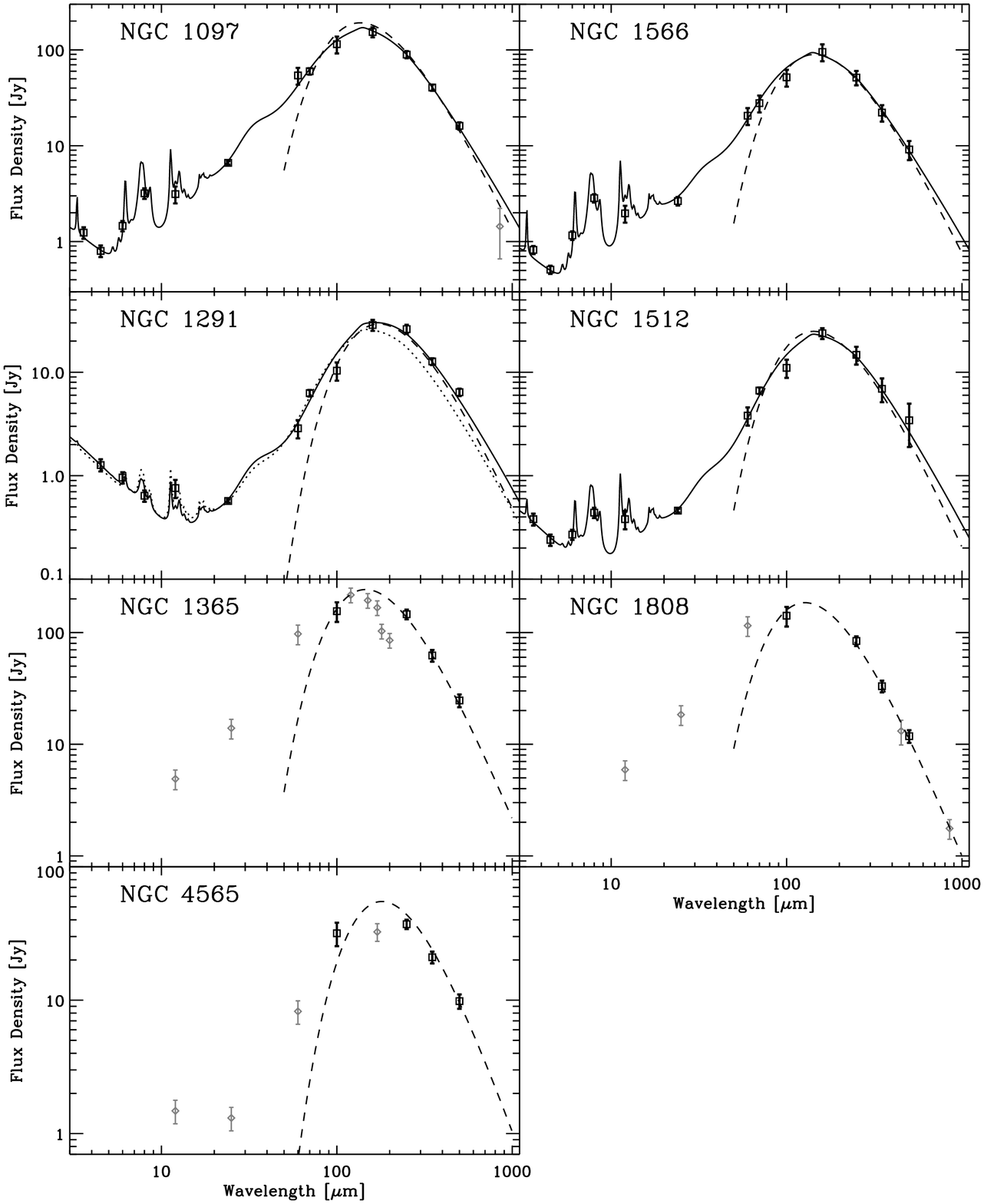}
\caption{
Spatially integrated SEDs for the seven galaxies in our sample.  Data 
from BLAST, MIPS, IRAC, and \IRAS\ used in this analysis are shown as black
squares.  Other data, which are not used in this analysis, from \IRAS\ for
NGC~1365, NGC~1808, and NGC~4565, SCUBA for NGC~1097 \citep{dale2005}, and
NGC~1808 \citep{stevens2005}, and \ISO\ for NGC~1365, and NGC~4565
\citep{spinoglio2002,stickel2004} are the lighter diamonds.
The aperture corrections calculated in \S\,\ref{sec:multi} have been applied
to the SCUBA measurements of NGC~1808.
%The SCUBA 850\,\um\ measurement of NGC~1097 \citep{dale2005} is also shown, but
%not included in this analysis.r
The solid curves are the best-fit
physical dust models of \citet{draine2007li} ({\sl DL07}; see
\S\,\ref{sec:dl07}).  The dotted curve in the panel for
NGC~1291 is the best-fit DL07 model subject to the constraint $\Umin\ge0.7$.
The dashed curves are the best-fit single-component modified blackbody model
(\S\,\ref{sec:globsed}).  Parameter values for these fits are given in
Tables~\ref{tab:fluxen} and~\ref{tab:dl07}.
\label{fig:glosed1}
}
\end{figure*}

The best-fit models for the galaxies to the extended \Spitzer-\IRAS-BLAST data
set are plotted in Fig.~\ref{fig:glosed1}.  This figure also plots the
spatially integrated \Spitzer\ and \IRAS, BLAST, and SCUBA measurements for
these galaxies.

%\begin{figure*}
%\centering
%\includegraphics[width=6in]{f9.eps}
%\caption{
%Spatially integrated SEDs for the three galaxies not in the SINGS sample.
%Modified blackbody fits to BLAST plus \IRAS\ 100\,\micron\ are shown.
%Other data, which were not included in the fit, from SCUBA \citep{stevens2005},
%\ISO\ \citep{spinoglio2002,stickel2004} and \IRAS\
%are the lighter diamonds.  Masses and
%temperatures from these fits are presented in Table~\ref{tab:fluxen}.
%\label{fig:glosed2}
%}
%\end{figure*}

The model parameters are
tabulated in Table~\ref{tab:dl07}.  We also tabulate the dust-weighted mean
starlight intensity scale factor
\begin{equation}
\left<U\right>=\left[ (1-\gamma) \Umin + \gamma \frac{\ln(\Umax/\Umin)}{\Umin^{-1} - \Umax^{-1}} \right ]
\end{equation}
and the dust luminosity calculated from the model fit.

\subsection{Modified Blackbody SEDs}

In addition to the DL07 models, we also fit BLAST plus available MIPS
70 and 160\,\um\ and \IRAS\ 100\,\micron\ data to a single-component modified
blackbody SED:
\begin{equation}
F_{\nu,\rm model}=\frac{\Mdust \kappa}{D^2} \left( \nu \over \nu_0 \right)^\beta\!\!\!B_\nu(T)\rm ,
\label{eq:gb}
\end{equation}
where the dust emissivity index, $\beta$, is fixed to 2, and $\nu_0$ to
1.2\,THz${}=c/$250\,\micron.  An uncertainty of $\pm0.3$~Mpc is assumed for all
galaxy distances.  This simplified SED model is needed in the absence of
\Spitzer\ data, where the DL07 models are insufficiently constrained.  We also
use this model in our subgalactic SED fits (\S\,\ref{ppsf}) due to its
computational simplicity.

The dust mass absorption co-efficient at $\nu_0$,
$\kappa$, is not well determined, and depends on the object investigated
(see \citealt{netterfield2009} for a discussion).  For each BLAST/SINGS galaxy,
we calculate a value for $\kappa$ by comparing the DL07 fit to the fit found for
the
spatially integrated modified blackbody SED (\S\,\ref{sec:globsed}).  The values
of $\kappa$ for each galaxy are summarized in Table~\ref{tab:dl07}.  These
values are used in the calculation of the column density maps presented in
Fig.~\ref{fig:allmass}.
For the galaxies without \Spitzer\ observations, we use the
mean value, $\kappa_{250}=0.29\pm0.03$\,m$^2$\,kg$^{-1}$, to calculate dust mass
from the modified blackbody fits.  Assuming a $\kappa \propto \nu^2$ evolution,
this corresponds to $k_{850}=0.025$\,m$^2$\,kg$^{-1}$, smaller than the
$k_{850}=0.07$\,m$^2$\,kg$^{-1}$ value calculated by \citet{james2002}.  It is
also lower than the 250\,\um\ values of \citet{draine1984} and
\citet{casey1991}.

For each fit,
a brute-force grid search is performed to simultaneously fit \Mdust\ and $T$
to the data.  The temperature, $T$, is sampled at 1000 uniformly-spaced
points between 5\,K and 30\,K.  This range was determined by first letting
temperatures vary over a much wider range (1\,K--100\,K) and then discarding
those portions of the parameter space discovered to be irrelevant to the fitting
procedure.  Because the dust mass, \Mdust, may
vary by several orders of magnitude, the sampling limits for the dust mass are
automatically adjusted as necessary for the fit, and sampled uniformly at 1000
points.  As with the DL07 models, goodness-of-fit is assessed using a
$\chi^2$ test (\S\,\ref{sec:chi2}).  The marginalized likelihood is used to
determine medians and 68\% confidence intervals for the two fit parameters.

\subsection{Spatially Integrated SED Fits}
\label{sec:globsed}
For the three galaxies without \Spitzer\ data, we fit Equation~\ref{eq:gb} to 
BLAST plus \IRAS\ 100\,\um.  The IRAS and BLAST measurements used to constrain
the fits, and the results from these fits are presented in
Fig.~\ref{fig:glosed1}, which also plots additional data from \IRAS, \ISO\
and SCUBA, not used in this analysis.

For the four BLAST/SINGS galaxies, we also fit Equation~\ref{eq:gb} to BLAST,
plus 160, 100, and 70\,\um\ data from MIPS or \IRAS, to compare with the results
of the DL07 models.
These fits are plotted in Fig.~\ref{fig:glosed1} as dashed curves.
We note reasonably good agreement between the two models for all galaxies near
the thermal peak.  On the Wien side of the peak, as expected, the
single-component modified blackbody
quickly falls off in comparison to the DL07 model fit.  On the Raleigh-Jeans
side, due to the adoption of an emissivity index $\beta=2$,
the modified blackbody also falls off faster than the DL07 model fit.
For NGC~1097 and NGC~1291, the modified blackbody fit
underpredicts the 500\,\um\
flux density by more than 1-$\sigma$.

Dust temperatures from these fits for all galaxies in the sample are listed in
Table~\ref{tab:fluxen}, as are dust masses for the three galaxies without
\Spitzer\ observations.

\subsection{Subgalactic SED Fits}
\label{ppsf}

Because the MIPS beams at 70\,\um\ and 160\,\um\ are of comparable size to the
BLAST beams, we also investigate the dust properties on sub-galactic scales
for the four BLAST/SINGS galaxies.
To do this, the BLAST maps produced for
the four galaxies were made on the same 9\arcsec\ pixel grid as the MIPS
160\,\um\ maps.
The 70\,\um\ MIPS images, which have a compatibly aligned 4\farcs5
pixel grid, are rebinned to 9\arcsec\ to put them on the same grid.  All five
maps are then convolved (smoothed) to the resolution of the 500\,\um\ data
(60\arcsec) and our modified
blackbody template (Equation~\ref{eq:gb}) is fit to the data at each pixel in
the map.

The central twelve
pixels of NGC~1097 have 70\,\micron\ surface brightness
in excess of 200\,MJy\,sr$^{-1}$.  For
these twelve pixels, the 70\,\micron\ data are excluded, due to possible
non-linearities\rlap,\footnote{\myurl{http://data.spitzer.caltech.edu/popular/sings/ 20070410\_enhanced\_v1/Documents/sings\_fifth\_delivery\_v2.pdf}}
and the model is fit to four photometric
points only.
%This results in these pixels having fits with larger uncertanties than
%neighbouring pixels.
% Fig.~\ref{fig:70um1097} presents an example of this.

%\begin{figure}
%\centering
%\rotatebox{270}{
%\includegraphics[width=3in]{figs/70um1097.eps}
%}
%\caption{
%SED data and fits for two adjacent pixels near the centre of NGC~1097.  The
%light grey line and points are for a pixel for which the 70\,\micron\ flux
%is above 200\,MJy\,sr$^{-1}$, and therefore ignored (but shown here without
%errors).  The black curve and points are for an adjacent pixel where all
%five points are fit to.  Dashed error contours, were the errors Gaussian and
%uncorrelated, are shown.  Without the 70\,\micron\ point, the remaining data
%poorly constrain the SED.\label{fig:70um1097}
%}
%\end{figure}

\begin{figure*}
\centering
\includegraphics[width=6in]{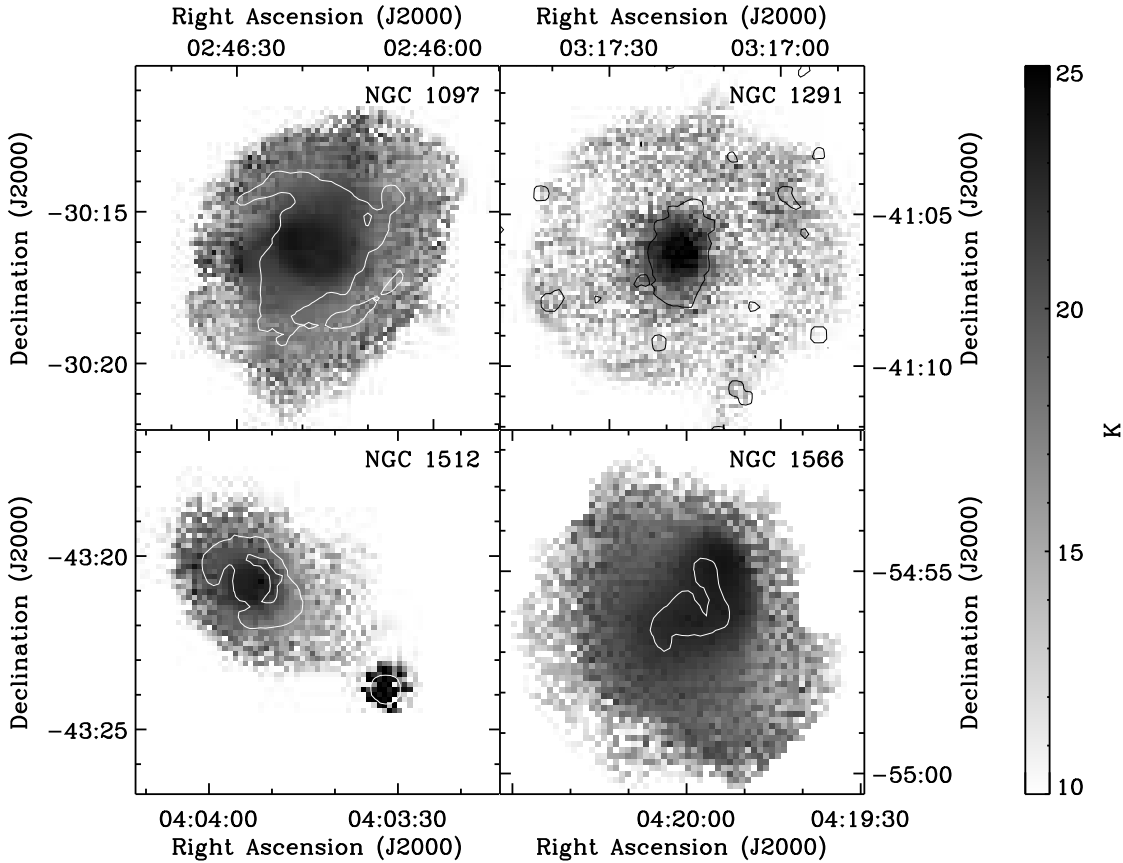}
\caption{
Dust temperature maps based on BLAST and MIPS observations for the four
BLAST/SINGS galaxies.
Only those pixels whose 250\,\micron\ flux density is inconsistent with
zero at the 1-$\sigma$ level are plotted.  Contours from the 24\,\micron\ MIPS
maps, smoothed with a 18\arcsec$\times$18\arcsec\ boxcar window function, are
over-plotted for
comparison.  For NGC~1097 and NGC~1512, the 0.3\,MJy\,sr$^{-1}$ 24-\um\ contour
is plotted.  For NGC~1291, the 0.1\,MJy\,sr$^{-1}$ contour is used, and for
NGC~1566, the 5\,MJy\,sr$^{-1}$ contour is plotted.
Mean temperature errors in the high SNR portions of the maps are $\pm$0.8\,K,
$\pm$2.5\,K, $\pm$1.1\,K and $\pm$0.8\,K for NGC~1097, NGC~1291, NGC~1512, and
NGC~1566, respectively.
\label{fig:alltemp}
}
\end{figure*}

\begin{figure*}
\centering
\includegraphics[width=6in]{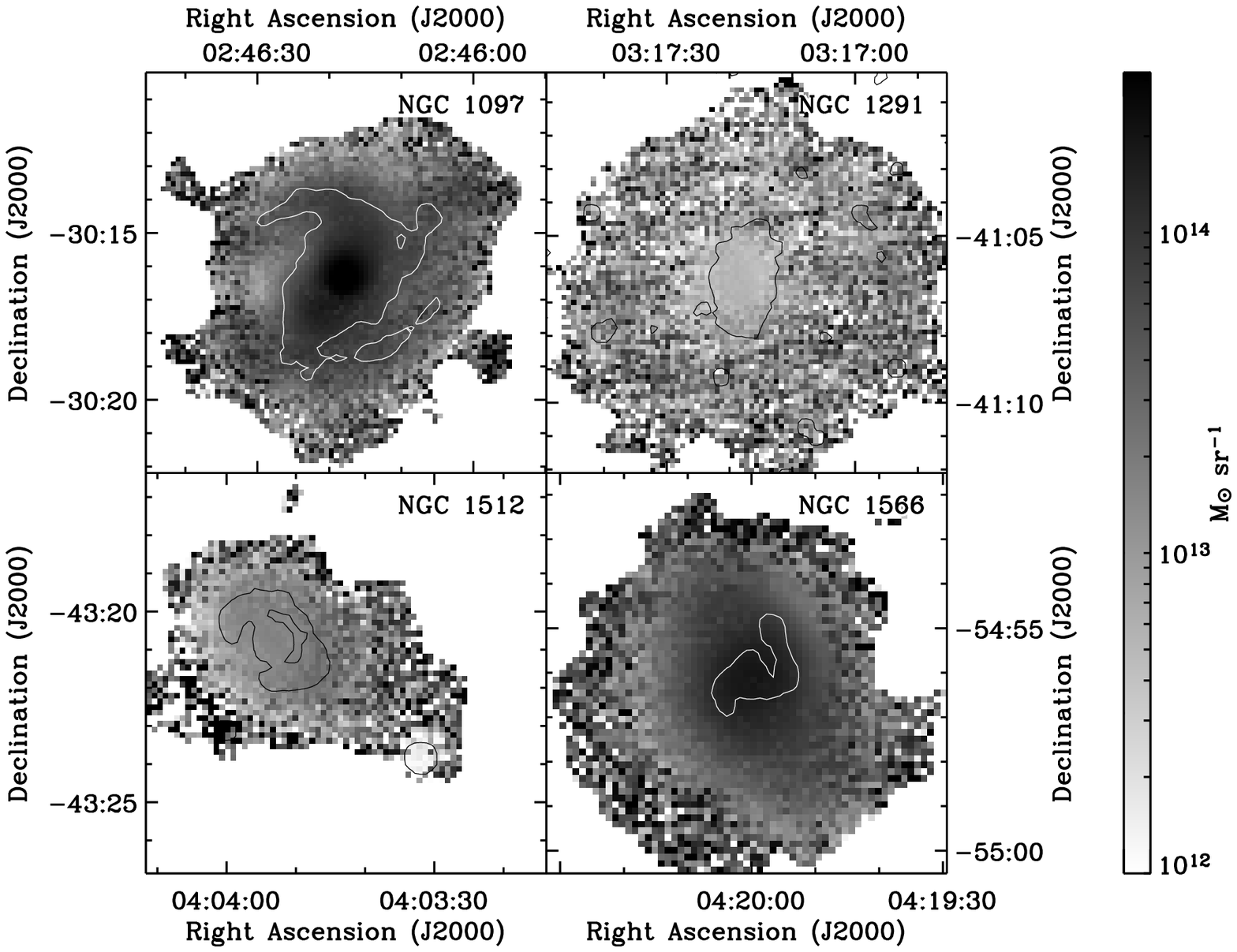}
\caption{
Dust column density maps based on BLAST and MIPS observations for the four
BLAST/SINGS galaxies.
Only those pixels whose 250\,\micron\ flux density is inconsistent with
zero at the 1-$\sigma$ level are plotted.  Contours plotted are the same as in
Fig.~\ref{fig:alltemp}.
Mean dust column density errors in the high SNR portions of the maps are
$\pm$2.1$\times10^{13}$\,$M_\odot$\,sr$^{-1}$,
$\pm$1.6$\times10^{13}$\,$M_\odot$\,sr$^{-1}$,
$\pm$0.9$\times10^{13}$\,$M_\odot$\,sr$^{-1}$, and
$\pm$2.1$\times10^{13}$\,$M_\odot$\,sr$^{-1}$
for NGC~1097, NGC~1291, NGC~1512, and NGC~1566, respectively.
\label{fig:allmass}
}
\end{figure*}

Resultant temperature and column density maps for the four galaxies are
presented
in Figs.~\ref{fig:alltemp} and~\ref{fig:allmass}.  NGC~1510, the companion
galaxy to NGC~1512, appears quite prominently in these maps as seemingly
hotter ($>$25~K) than NGC~1512.
This is not believed to be a real effect, and is likely due to poor performance
by the fitting routine on the data for NGC~1510, for which $\beta=2$
is probably not an appropriate choice.
%(see Fig.~\ref{fig:ngc1510})

\begin{figure}
\centering
\includegraphics[width=3in]{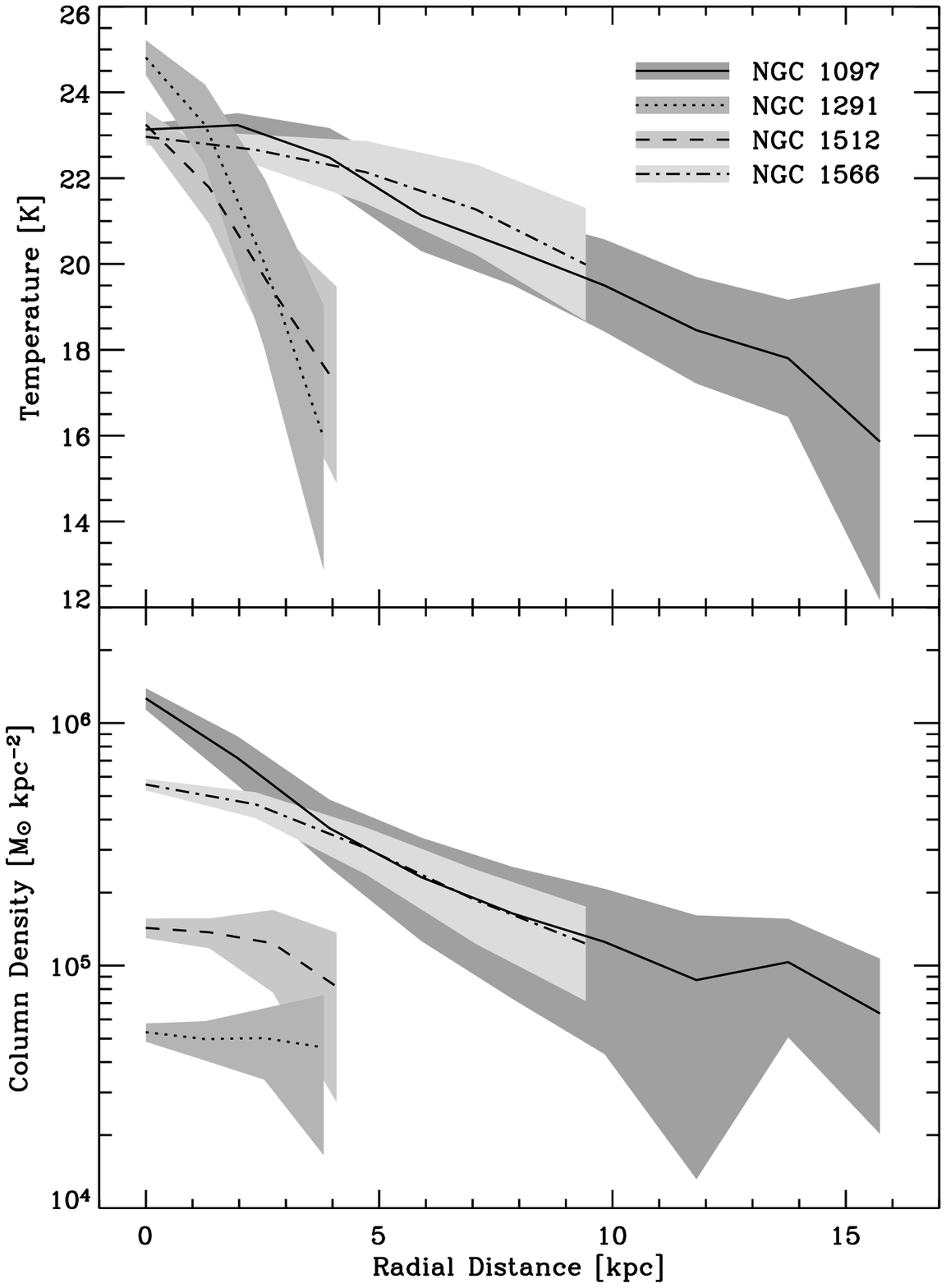}
\caption{
Mean radial dust temperature (top) and dust column density (bottom) for the four
BLAST/SINGS galaxies,
with 1-$\sigma$ uncertainty bands.  Resolution elements at 500\,\micron\
are approximately 4.9, 5.2, 3.0, and 2.8\,kpc for NGC~1097, NGC~1291, NGC~1512,
and NGC~1566, respectively.
\label{fig:rad}
}
\end{figure}

Radially-averaged temperature
and column density profiles are plotted in Fig.~\ref{fig:rad}.  We note
a general decrease in both column density and dust temperature with increasing
radius.  Of the four galaxies, we notice two subgroups of two galaxies each.
The brighter NGC~1097 and NGC~1566, for which BLAST detects high intensity
emission continuously from the core to disk of the galaxy, have relatively
flat temperature profiles.  NGC~1291 and NGC~1512, which BLAST detects with
a more isolated core, fall off more steeply.  We also potentially detect the
1.5~kpc star-forming ring in NGC~1097, which causes the kink in the temperature
profile.

%\begin{figure}
%\centering
%\rotatebox{270}{
%\includegraphics[width=3in]{figs/ngc1510.eps}
%}
%\caption{
%Data and fit with $\beta=2$ for a single pixel near the centre of NGC~1510
%demonstrating the poor fit for this galaxy.  Pixels around NGC~1510 prefer a
%lower $\beta$.
%\label{fig:ngc1510}
%}
%\end{figure}

\subsection{$\chi^2$ Calculation}
\label{sec:chi2}

We use $\chi^2$ minimization to assess goodness of fit for both the DL07 models
and the modified blackbody model.  The BLAST calibration uncertainties are
highly
correlated between observational bands \citep{truch2008,truch2009}.  Given
observational data, $s_{\rm d}$, and band-convolved model predictions,
$\tilde s$, we compute
\begin{equation}
\chi^2 \equiv \left(\tilde s - s_{\rm d}\right)^{\rm T}
C^{-1} \left(\tilde s - s_{\rm d}\right)\rm ,
\end{equation}
where $C^{-1}$ is the inverse data covariance matrix, and $(\cdot)^{\rm T}$
denotes the transpose.
Given the calibration uncertainty correlation matrix, $\rho$, the data
covariance matrix is:
\begin{equation}
C_{ij} = \delta_{ij} \left[ \left(\sigma^2_{\rm d}\right)_i +
\left(\sigma^2_{\rm m}\right)_i \right] +
\rho_{ij} \left(\sigma_{\rm c}\right)_i
\left(\sigma_{\rm c}\right)_j\rm ,
\end{equation}
where $\sigma_{\rm d}$ is the measurement uncertainty, $\sigma_{\rm c}$ the
calibration uncertainty, and $\sigma_{\rm m}$ an uncertainty associated with
the model.  For the DL07 models, in order to compare with their results, we
follow Dr07 and adopt $\sigma_{\rm m}=0.1
s_{\rm d}$.  For the modified blackbody model, we use $\sigma_{\rm m}=0$.
We assume that \IRAS\ and \Spitzer\ uncertainties are uncorrelated between
bands, and set $\rho_{ij} = \delta_{ij}$ for non-BLAST measurements.

For the
BLAST measurements, we set $\rho$ to the value of the Pearson correlation
matrices computed as outlined in \citet{truch2008, truch2009}.
For both the calibration
uncertainty ($\sigma_{\rm c}$) and the calibration uncertainty correlation
matrix ($\rho$), we add to the values given in those papers an
additional source of uncertainty associated with our ability to measure the
BLAST band passes.
We estimate the error in the bandpass measurements to be at most 5\%.
We additionally assume that this uncertainty is uncorrelated between bands.
%although this assumption is debatable.

The magnitude of the effect of bandpass measurement error on the BLAST
calibration uncertainties depends on the spectrum of the object observed.
A source with the same spectrum as our calibrators (e.g., for BLAST06,
VY Canis Majoris, for which \citet{truch2009} find a best-fit single-component
modified blackbody with
$T=346\pm19$\,K and $\beta=0.55\pm0.05$) is unaffected by bandpass errors,
but for a source with a significantly different spectrum, the effect can be
non-negligible, especially in the high SNR case.  This is the case in this
study, and we have adjusted our calibration uncertainties accordingly.

Because the underlying spectrum of the galaxies is unknown, we use the
modified blackbody model as a proxy.  The calculated
calibration uncertainties affect the best fit of the model, so this is an
iterative process: we first fit the model using a guess for the calibration
uncertainties, and then update the uncertainties based on the best fit.  This
process is iterated until convergence is achieved.  In general only one
iteration after the initial guess is required for convergence.

For this procedure, we have calculated calibration uncertainties and uncertainty
correlations for
various \hbox{$\beta=2$} modified blackbody models with temperatures in the range
15--25\,K.  The resulting calibration uncertainties vary most with choice of
model at 250\,\um\ (12--14\% for BLAST06), with smaller variations at 350\,\um\
(10.0--10.5\%) and 500\,\um\ (8.00--8.25\%).  The uncertainty correlations
calculated show similar behavior, varying between 0.80 and 0.97 for the
250--350\,\um\ correlation but only 0.91--0.92 for the 350--500\,\um\
correlation.
%half degree intervals for the calibrations of both BLAST flights.  These values
%are reported in Tables~\ref{tab:cal05} and~\ref{tab:cal06}.

When reporting goodness of fit, we report $\chi^2$ per degree of freedom:
\begin{equation}
\chi^2_r \equiv \frac1{N_{\rm b} - N_{\rm m}}\chi^2
\end{equation}
where $N_{\rm b}$ is the number of observational bands considered in the fit,
and $N_{\rm m}$
the number of model parameters (five for the DL07 models and two for the
modified blackbody model).  The number of bands ranges from four (3~BLAST +
100\,\um\ \IRAS) for modified blackbody fits in the absence of \Spitzer\ data,
to thirteen (4~IRAC, 3~\IRAS, 3~MIPS, 3~BLAST) when fitting the DL07 models.

\section{Discussion}

\subsection{Dust Models}

Of the four galaxies observed by both BLAST and \Spitzer, the DL07 fits we
find for all but NGC~1291 are consistent with the best-fit models determined
without BLAST data in Dr07.  For NGC~1291 we find a dust mass approximately
two times larger than Dr07.  This discrepancy is due to the restriction
$\Umin \ge 0.7$ imposed by Dr07 in the absence of SCUBA data.  We fore-go this
restriction for our fit of NGC~1291, and find a best-fit $\Umin=0.2$.
Constraining $\Umin\ge0.7$ results in a fit consistent with that of Dr07
(\qpah=2.8\%, \Umin=0.7, log(\Mdust/$M_\odot$)=7.36, $\gamma=1.00$\%), but a
poor fit to BLAST data ($\chi^2_r=2.59$). This fit is plotted in
Fig.~\ref{fig:glosed1} for comparison.

%\subsection{Galaxy Colors}
%\label{sec:colcol}
%
%The high SNR BLAST measurements in
%Table~\ref{tab:fluxen} strongly constrains the shape of the galactic SEDs.
%Using this shape information, we can attempt to determine $T$ and $\beta$
%for the single-component modified blackbody model (Equation~\ref{eq:gb})
%directly.  All seven galaxies in our sample are shown on a color-color plot in
%Fig.~\ref{fig:colcol}.
%Also plotted are various isotherms and iso-$\beta$ contours.  This
%single-component modified blackbody
%model is clearly inconsistent, regardless of choice of $\beta$,
%with at least some of these high signal-to-noise
%data, which is not surprising given the simplicity of the model.

%\begin{figure}
%\centering
%\includegraphics[width=3in]{f13.eps}
%\caption{
%Color-color plot of the seven BLAST galaxies.  Crosses indicate 1-$\sigma$
%errors, assuming the BLAST calibration uncertainty is 100\% correlated.
%Contours of both constant $T$ varying $\beta$ and constant $\beta$ varying
%$T$ from the modified blackbody model (equation~\ref{eq:gb}) are also
%shown.  The model is clearly inconsistent with some of the data.
%\label{fig:colcol}
%}
%\end{figure}

The dust temperatures found in this study agree well with the 15--25\,K
temperatures found for the cold dust component in other studies of nearby
galaxies
\citep[e.g.][]{alton1998,braine1999,
dunne2001,xilouris2004,vlahakis2005,willmer2009}.

\subsection{Fraction of Core to Total Emission}

BLAST's unique ability to spatially resolve the submillimeter flux of these
nearby galaxies allows us to investigate the
fraction of submillimeter flux which arises from the extended
disks compared to their central core and circumnuclear regions.
%sample fully, along with its resolution, allow us to investigate the fraction of
%submillimeter flux which arises from the central core of these galaxies.
Although BLAST's resolution is insufficient to resolve the central AGN of these
galaxies (where present), we can still put limits on the contribution made to
the submillimeter flux by the nuclear accretion and star formation activity
towards their circumnuclear regions (typically within a few
kpc of the nucleus).
%active galactic nuclei.  This is important for the
%determination of star formation rates from submillimeter observation of distant
%galaxies \citep[e.g.,][]{devlin2009}.

For each galaxy, except the edge-on NGC~4565, we find the 3~db radial flux
contour (that is,
the radius at which the radially averaged flux has dropped to half of the
peak flux) at 250\,\um\ and use this radius to define the core of the
galaxy.  Fluxes in all three BLAST bands from this core region are presented in
Table~\ref{tab:agnfrac}.  Also in this table is the ``core fraction,'' the
fraction of the total BLAST measured flux which originates in the core of the
galaxy.

Most of the sample have core fractions in the range 7.5\% to 15\%.
NGC~1566 is a clear outlier, with roughly one
third of the submillimeter flux in the core of the galaxy.  The second
brightest known Seyfert galaxy \citep{devaucouleurs1973}, NGC~1566, has a core
whose submillimeter flux originates largely from the AGN.  The galaxy with the
smallest core fraction is NGC~1291, a galaxy with no known active nucleus.
Other than in NGC~1566, the bulk of the
submillimeter emission in the core of the galaxy could be explained by nuclear
star formation.

All galaxies show a decrease in core fraction with increasing wavelength, a
result of the core dust temperature being generally warmer than the dust in the
surrounding disk (see Fig.~\ref{fig:rad}).  Other than NGC~1566, most galaxies
have weak or no AGN, although BLAST is unable to separate AGN derived flux
from star formation in the central regions of the galaxies.  In the case of
NGC~1097, the unresolved 18\arcsec\ circumnuclear starburst ring
\cite[e.g.,][]{hummel1987,barth1995,quillen1995,kotilainen2000}, likely provides
a sizable contribution to the total core flux.

%In only one of our galaxies, NGC~1566, is the AGN clearly driving a significant
%fraction of the submillimeter detected flux.
BLAST's ability to measure the compact to extended flux
ratio of these galaxies is a result of BLAST's high sensitivity and resolution.
This indicates that current and future, higher-sensitivity and higher-resolution
FIR and submillimeter experiments, such as \Herschel, \SCUBATOO, {\it SPICA},
{\it JWST}, ALMA, LMT, and GBT, which will be able to spatially separate
non-thermal components from extended star formation emission, are important for
understanding the relationship between FIR-submillimeter luminosity and star
formation rates in both local and high-redshift galaxies.

%A simple model based on the mean flux of the disk of the galaxies
%and the mean flux of NGC~1291 indicates that approximately 5\% of the
%submillimeter luminosity of LIRGS and ULIRGS comes from star formation
%driven heating in the core.  In many galaxies, an excess of 2.5-10\% (and as
%high at 25\%) appears to come from AGN-driven flux.  The result leads to a
%significant overestimate of star formation in galaxies with AGN, unless this
%effect can be taken into account.

\subsection{Comparison of Resolved and Unresolved Mass and Luminosity Estimates}

Our resolved measurements of the mass distribution in these galaxies allow us to
investigate how mass estimates may be affected in more distant and smaller
galaxies that are unresolved.
For the four galaxies for which we have dust column density
maps (Fig.~\ref{fig:allmass}), we apply the same aperture used to calculate
our total flux measurements to the column density maps.  The masses contained in
these apertures, along with the aperture diameters,
are listed in Table~\ref{tab:mass}.

In all cases, the resolved mass estimate is significantly higher than the mass
estimate obtained by the DL07 fit to the unresolved total flux from the galaxy.
The ratio of resolved to unresolved mass estimates
ranges from $\sim$5, for NGC~1566, to
$\sim$19, for NGC~1291.  Furthermore, the mass ratio follows the behavior of
the radial temperature profile (Fig.~\ref{fig:rad}), with a larger temperature
gradient corresponding to a larger mass ratio.  This effect is independent of
the choice of $\kappa$, since both our resolved and unresolved mass estimates
use the same $\kappa$ to derive dust mass from the SED fit.

As a check, we performed a similar analysis to measure the difference between
resolved and unresolved luminosity estimates.  Using the same aperture, we
summed the 60--1000\,\um\ luminosity in the resolved maps, and compared this
to the 60--1000\,\um\ found from the modified blackbody fit to the galaxy as a
whole.  As expected, these
values agree with one another to within measurement uncertainty.

Clearly, while luminosity, and hence star-formation rate, is unaffected by
resolution,
accurate dust mass measurements rely on resolved observations.
%Because mass is a
%strong function of temperature, unresolved observations of galaxies tend to hide
%cool dust, when this cool dust is confused with a warmer, less massive core.
Since luminosity, and therefore observed flux density, is a strong
function of temperature, there is a potential for unresolved observations
of galaxies to hide cool dust. For the four galaxies we investigate, a
disproportionate fraction of their total SEDs is produced by their relatively
warmer, less massive cores. A single temperature modified blackbody fit to
such data tends to infer a warmer temperature and shallower (i.e. smaller)
value of $\beta$, and hence lower, incorrect, dust mass.  Our data are
consistent with the model of \citet{dunne2001}, who use a two-component $\beta=2$ modified
blackbody SED to separate the cool (15--25\,K) dust component from warmer
($T>30$\,K) emission in the SCUBA Local Universe Galaxy Survey (SLUGS).

The upcoming
higher-resolution experiments should be able to better estimate the quantity of
dust in distant galaxies.

\section{Summary}

BLAST made resolved observations of seven nearby galaxies over the course of
two flights.  Four of these galaxies, NGC~1097, NGC~1291, NGC~1512, and
NGC~1566, have complementary \Spitzer\ observations as part of the SINGS survey.
For these four galaxies, we fit the models of \cite{draine2007li} to BLAST,
\Spitzer, and \IRAS\ observations.  Best-fit parameters are tabulated in
Table~\ref{tab:dl07}.

The best-fit models for three of these four galaxies
are consistent with the best-fit models calculated in \cite{draine2007}
without BLAST data.  For the fourth galaxy, NGC~1291, we find a dust
mass roughly two times larger than the dust mass determined by
\cite{draine2007}.
We also calculate a value for the dust mass absorption coefficient at 250\,\um\
$\kappa=0.29\pm0.03$\,m$^2$\,kg$^{-1}$ by comparing the
Draine \& Li models with a modified blackbody model.

For these four galaxies, we also produce maps of dust column density and mean
dust temperature based on BLAST and MIPS observations fit to a single component
modified blackbody
model (Figs.~\ref{fig:alltemp} and~\ref{fig:allmass}), as well as radial
profiles of the same quantities (Fig.~\ref{fig:rad}).
For the remaining three galaxies observed by BLAST but not observed by \Spitzer,
NGC~1365, NGC~1808, and NGC~4565, we calculate spatially integrated dust
temperatures and dust masses by fitting a modified blackbody model to BLAST
and \IRAS\ data.  We find mean dust temperatures for our galaxy sample to be in the range 16--23~K.

We calculate the fraction of the submillimeter detected flux originating in the
core of the galaxy as a fraction of the total submillimeter emission from the
galaxy.  Although BLAST is unable to resolve the nucleus of these galaxies, the
ratio of compact to extended flux measured by BLAST puts an important upper
limit on the fraction of submillimeter radiation driven by a central AGN.

Only one of our sample, NGC~1365, the second brightest known Seyfert
galaxy, has a significant fraction of its flux ($>$30\%) emanating from the
core of the galaxy.  In the remainder the core fraction is small ($<$15\%) and
may be accountable primarily to nuclear star formation.
%We find in the AGN-quiet NGC~1291, approximately 5\% of the total BLAST detected
%flux arises in the core.  In much of our sample, the core fraction is 8\% or
%higher.  In the extreme case of NGC~1365, the second brightest known Seyfert
%galaxy, a third of the submillimeter flux originates in the core.
In all cases,
the core fraction increases with decreasing wavelength, indicating that dust
in the core of these galaxies is warmer than the dust in the surrounding disk,
agreeing with the radial temperature profile of Fig.~\ref{fig:rad}.

We compare the total dust mass in our column density maps to the dust mass
estimate based on the observation of the galaxies as a whole (i.e., an
unresolved observation).  We find 5--19 times more dust mass in the resolved
maps than the unresolved estimate, a result of the unresolved measurements hiding
cool dust.  We perform a similar analysis with luminosity and
find no discrepancy between the resolved and unresolved 60--1000\,\um\
luminosity estimates.

Finally,
for NGC~4565, we calculate a scale length of 118\arcsec, 156\arcsec, 142\arcsec\
at 250\,\um, 350\,\um, and 500\,\um, respectively, in good agreement with
measurements at other wavelengths.

\acknowledgments
The BLAST collaboration acknowledges the support of NASA through grant numbers
NAG5-12785, NAG5-13301, and NNGO-6GI11G, the Canadian Space Agency (CSA),
Canada's Natural Sciences and Engineering Research Council (NSERC), and the UK
Particle Physics \& Astronomy Research Council (PPARC).
We would also like to thank the Columbia Scientific Balloon Facility (CSBF)
staff for their outstanding work.
LO acknowledges partial support by the Puerto Rico Space Grant Consortium and
by the Fondo Istitucional para la Investigacion of the University of Puerto
Rico.
CBN acknowledges support from the Canadian Institute for Advanced Research.
This research has been enabled by the use of WestGrid computing resources.

This work is based in part on observations made with the
{\it Spitzer Space Telescope\/},
%which is operated by the Jet Propulsion Laboratory, California
%    Institute of Technology under a contract with NASA.
%
%This research has
and has also
made use of the NASA/IPAC Extragalactic Database (NED),
%which is
both of which are 
operated by the Jet Propulsion Laboratory, California Institute of Technology,
under contracts with the National Aeronautics and Space Administration.
This research also made use of the SIMBAD database, operated at the Centre de
Don{\'e}es astronomiques de Strasbourg (CDS), Strasbourg, France.

\bibliographystyle{apj}
\bibliography{ms}

\clearpage

\begin{deluxetable}{lrrrrc}
\tablecaption{Physical parameters of BLAST nearby galaxies\label{tab:physparm}}
\tablewidth{0pt}
\tablehead{
\multirow{2}{*}{Galaxy} & \colhead{R.A.} & \colhead{Dec.} & \colhead{Dist.}
& \multirow{2}{*}{Type\tablenotemark{{(a)}}}
& \multirow{2}{*}{Nucleus} \\
& \colhead{(J2000)} & \colhead{(J2000)} & \colhead{[Mpc]}
}
\startdata
NGC~1097 & 2\hour46\minute19\seconddot0 & $-$30\dg16\arcmin30\arcsec
& 16.9\tablenotemark{{(b)}}\pho & SB(s)b
& Sy1
\\ %.SBS3..
NGC~1291 & 3\hour17\minute18\seconddot6 & $-$41\dg06\arcmin29\arcsec
& 9.7\tablenotemark{{(b)}}\pho & (R)SB(s)0/a
& \nodata
\\ % RSBS0..
NGC~1365 & 3\hour33\minute36\seconddot4 & $-$36\dg08\arcmin25\arcsec
& 18.4\tablenotemark{{(c)}}\pho & SB(s)b
& Sy2
\\ % .SBS3..
NGC~1512 & 4\hour03\minute54\seconddot3 & $-$43\dg20\arcmin56\arcsec
& 10.4\tablenotemark{{(b)}}\pho & SB(r)a
& \nodata
\\ % .SBR1..
NGC~1566 & 4\hour20\minute00\seconddot4 & $-$54\dg56\arcmin16\arcsec
& 18.0\tablenotemark{{(b)}}\pho & SAB(s)bc
& Sy1
\\ % .SXS4..
NGC~1808 & 5\hour07\minute42\seconddot3 & $-$37\dg30\arcmin47\arcsec
& 10.9\tablenotemark{{(d)}}\pho & (R)SAB(s)a
& Sy2
\\ % RSXS1..
NGC~4565 & 12\hour36\minute20\seconddot8 & $+$25\dg59\arcmin16\arcsec
& 12.6\tablenotemark{{(e)}}\pho & SA(s)b
& Sy2
\\ % .SAS3$/
\enddata

\tablenotetext{a}{RC3 morphological type from \citet{devaucouleurs1991}}
\tablenotetext{b}{\citet{kennicutt2003}}
\tablenotetext{c}{\citet{komossa1998}}
\tablenotetext{d}{\citet{jimenez-bailon2005}}
\tablenotetext{e}{\citet{lawrence1999}}
\end{deluxetable}

\begin{deluxetable}{lrrrrc}
\tablecaption{Observed flux densities and derived quantities for
BLAST nearby galaxies\label{tab:fluxen}}
\tablewidth{0pt}
\tablehead{
\multirow{2}{*}{Galaxy} & \colhead{250\,\um} & \colhead{350\,\um}
& \colhead{500\,\um}
& \colhead{$\Tdust$}
& \multirow{2}{*}{$\log\left(\Mdust\over{M_\odot}\right)$}
\\
& \colhead{[Jy]} & \colhead{[Jy]} & \colhead{[Jy]} & \colhead{[K]}
}
\startdata
NGC~1097 &  89.4\plm0.2\plm\pho7.7 & 40.5\plm0.1\plm3.0
& 16.2\plm0.1\plm1.4 & \tempsym{21.3}{0.3}
& %\massasym{8.37}{0.12}{0.16}
\nodata\tablenotemark{{(a)}}
\\[0.4em]
NGC~1291 & 26.1\plm0.2\plm\pho2.6 & 12.7\plm0.1\plm1.0
& 6.4\plm0.1\plm0.5 & \tempsym{16.6}{0.4}
& %\massasym{7.51}{0.15}{0.21}
\nodata\tablenotemark{{(a)}}
\\[0.4em]
NGC~1365 & 145.8\plm0.3\plm12.9 & 62.3\plm0.2\plm4.6
& 24.7\plm0.1\plm2.1 & \tempsym{19.9}{0.4}
& \massasym{8.72}{0.08}{0.10}
\\[0.4em]
NGC~1512 & 14.7\plm0.1\plm\pho1.3 & 6.9\plm0.1\plm0.5
& 3.4\plm0.1\plm0.3 & \tempsym{20.3}{0.3}
& %\massasym{7.25}{0.14}{0.19}
\nodata\tablenotemark{{(a)}}
\\[0.4em]
NGC~1566 & 51.5\plm0.2\plm\pho4.6 &  22.2\plm0.1\plm1.7
& 9.1\plm0.1\plm0.8 & \tempsym{20.1}{0.4}
& %\massasym{8.30}{0.13}{0.17}
\nodata\tablenotemark{{(a)}}
\\[0.4em]
NGC~1808 & 84.0\plm0.1\plm\pho7.1 &  33.1\plm0.1\plm2.5
& 11.8\plm0.1\plm1.0 & \tempsym{22.8}{0.5}
& \massasym{7.85}{0.09}{0.11}
\\[0.4em]
NGC~4565 & 37.2\plm0.5\plm\pho4.5 & 21.0\plm0.3\plm2.1
& \pho9.8\plm0.3\plm0.8 & \tempasym{16.1}{0.4}{0.5}
& \massasym{8.21}{0.09}{0.11}
\enddata
\tablecomments{Flux uncertainties reported are first the measurement error
($\sigma_{\rm d}$), and then the calibration error ($\sigma_{\rm c}$).
Temperatures and dust masses are extracted from the
modified blackbody fits presented in Fig.~\ref{fig:glosed1}.
}
\tablenotetext{a}{This modified blackbody fit is used to calculate $\kappa$ from
the DL07 model fit, so no independent mass is determined.}
\end{deluxetable}

\begin{deluxetable}{lcccccccc}
\tablecaption{DL07 model parameters for BLAST nearby galaxies\label{tab:dl07}}
\tablewidth{0pt}
\tablehead{
\multirow{2}{*}{Galaxy}
& \multirow{2}{*}{$\log\left(\Mdust\over{M_\odot}\right)$}
& \multirow{2}{*}{$\log\left(\Ldust\over{L_\odot}\right)$}
& $q_{\rm PAH}$
& \multirow{2}{*}{$\left<U\right>$}
& \multirow{2}{*}{$U_{\rm min}$}
& $\gamma$
& \multirow{2}{*}{$\chi^2_r$}
& $\kappa$\tablenotemark{{(a)}}
\\
& & & \% & & & \% & & m$^2$\,kg$^{-1}$
}
\startdata
%NGC~1097 & 8.37 &    10.81 & 2.9 & 2.02 & 1.5 & 2.80 & 0.18 & 0.297 \\
NGC~1097 & 8.37 &    10.80 & 3.0 & 2.00 & 1.5 & 2.80 & 0.58 & 0.299 \\
NGC~1291 & 7.62 & \pho9.42 & 1.6 & 0.46 & 0.4 & 1.10 & 3.47 & 0.290 \\
NGC~1291\tablenotemark{{(b)}} &
7.43 & \pho9.45 & 3.1 & 0.77 & 0.7 & 0.70 & 5.66 & \nodata \\
NGC~1512 & 7.25 & \pho9.42 & 3.2 & 1.10 & 1.0 & 0.80 & 0.87 & 0.247 \\
NGC~1566 & 8.21 &    10.58 & 4.6 & 1.74 & 1.5 & 1.30 & 1.13 & 0.305 
\enddata
\tablenotetext{a}{at 250\,\um}
\tablenotetext{b}{with constraint $U_{\rm min}\ge0.7$}
\end{deluxetable}

\begin{deluxetable}{lcccrr}
\tablewidth{0pt}
\tablecaption{Core flux density
fractions for BLAST nearby galaxies\label{tab:agnfrac}}
\tablehead{
  & \multicolumn{2}{c}{Core} & \multirow{2}{*}{Band} & \colhead{Core}
  & \colhead{Core Flux}
  \\
    \colhead{Galaxy} & \multicolumn{2}{c}{Radius} && \colhead{Flux}
  & \colhead{Fraction}
  \\
    & \colhead{[\arcmin]} & \colhead{[kpc]} & \colhead{[\um]}
  & \colhead{[Jy]} & \colhead{[\%]}
}
\startdata
NGC~1097 & 0.37 & 1.8
  & 250 & 13.35\plm0.03 & 14.93\plm0.06 \\
&&& 350 &  4.97\plm0.02 & 12.26\plm0.07 \\
&&& 500 &  1.36\plm0.01 &  8.44\plm0.08 \\[0.4em]
NGC~1291 & 0.71 & 2.0
  & 250 &  1.41\plm0.02 &  5.42\plm0.10 \\
&&& 350 &  0.64\plm0.01 &  5.04\plm0.17 \\
&&& 500 &  0.23\plm0.01 &  3.58\plm0.17 \\[0.4em]
NGC~1365 & 0.29 & 1.6
  & 250 & 13.97\plm0.16 &  9.58\plm0.13 \\
&&& 350 &  5.38\plm0.10 &  8.62\plm0.19 \\
&&& 500 &  1.35\plm0.05 &  5.46\plm0.23 \\[0.4em]
NGC~1512 & 0.40 & 1.2
  & 250 &  1.51\plm0.01 & 10.26\plm0.08 \\
&&& 350 &  0.61\plm0.01 &  8.79\plm0.21 \\
&&& 500 &  0.14\plm0.01 &  3.43\plm0.06 \\[0.4em]
NGC~1566 & 0.98 & 5.1
  & 250 & 19.07\plm0.04 & 37.04\plm0.19 \\
&&& 350 &  7.72\plm0.04 & 34.78\plm0.37 \\
&&& 500 &  2.84\plm0.02 & 31.04\plm0.46 \\[0.4em]
NGC~1808 & 0.29 & 0.9
  & 250 & 10.86\plm0.01 & 12.94\plm0.03 \\
&&& 350 &  3.66\plm0.01 & 11.06\plm0.07 \\
&&& 500 &  0.92\plm0.01 &  7.79\plm0.05
\enddata
\tablecomments{Flux uncertainties reported are measurement errors
  ($\sigma_{\rm d}$) only.
}
\end{deluxetable}

\begin{deluxetable}{lcrrc}
\tablewidth{0pt}
\tablecaption{Resolved versus unresolved mass estimates
for BLAST nearby galaxies\label{tab:mass}}
\tablehead{
  & \colhead{Unresolved Mass\tablenotemark{{(a)}}}
  & \colhead{Resolved Mass}
  & \colhead{Ratio}
  & \colhead{Aperture}
  \\
    \colhead{Galaxy}
  & \multirow{2}{*}{$\log\left(M_{\rm u}\over{M_\odot}\right)$}
  & \multirow{2}{*}{$\log\left(M_{\rm r}\over{M_\odot}\right)$}
  & \multirow{2}{*}{$M_{\rm r}\over{M_{\rm u}}$}
  & Diameter
  \\
  &&&& \colhead{[\arcmin]}
}
\startdata
NGC~1097 & 8.37 & \massasym{9.17}{0.13}{0.14} & \tempasym{6.3}{2.2}{1.7} & 10.9
\\[0.4em]
NGC~1291 & 7.62 & \massasym{8.90}{0.17}{0.20} & \tempasym{19.1}{9.1}{7.0} & 13.2
\\[0.4em]
NGC~1512 & 7.25 & \massasym{8.38}{0.12}{0.14} & \tempasym{13.5}{4.3}{3.7} & 5.8
\\[0.4em]
NGC~1566 & 8.21 & \massasym{8.92}{0.09}{0.10} & \tempasym{5.1}{1.2}{1.1} & 6.8
\enddata
\tablenotetext{a}{DL07 model mass estimate from Table~\ref{tab:dl07}}
\end{deluxetable}

\end{document}